%% file: output.tex
\documentclass[fleqn,usenatbib]{mnras}

\usepackage[T1]{fontenc}

\DeclareRobustCommand{\VAN}[3]{#2}
\let\VANthebibliography\thebibliography
\def\thebibliography{\DeclareRobustCommand{\VAN}[3]{##3}\VANthebibliography}

\usepackage{graphicx}	
\usepackage{amsmath}	
\usepackage{amssymb}	
\usepackage{multirow}
\usepackage{xcolor}
\usepackage{newtxtext,newtxmath}

\title[]{The effect of phased recurrent units in the classification of multiple catalogs of astronomical lightcurves}

\author[C.Donoso Oliva]{C. Donoso-Oliva$^{1,3,2}$\thanks{E-mail: cridonoso@inf.udec.cl},
G. Cabrera-Vives$^{1,2}$,
P. Protopapas $^{3}$,
R. Carrasco-Davis $^{4,2,5}$,
P.A. Estevez $^{5, 2}$
\\\\
$^{1}$Dept. of Computer Science, Universidad De Concepcion\\
$^{2}$Millennium Institute of Astrophysics, Chile\\
$^{3}$Institute for Applied Computational Science, Harvard University\\
$^{4}$Gatsby Computational Neuroscience Unit, University College London\\
$^{5}$Dept. of Electrical Engineering, Universidad de Chile
}
\date{Accepted XXX. Received YYY; in original form ZZZ}

\pubyear{2015}

\begin{document}
\label{firstpage}
\pagerange{\pageref{firstpage}--\pageref{lastpage}}
\maketitle

\begin{abstract}
In the new era of very large telescopes, where data is crucial to expand scientific knowledge, we have witnessed many deep learning applications for the automatic classification of lightcurves. 
Recurrent neural networks (RNNs) are one of the models used for these applications, and the LSTM unit stands out for being an excellent choice for the representation of long time series. 
In general, RNNs assume observations at discrete times, which may not suit the irregular sampling of lightcurves. A traditional technique to address irregular sequences consists of adding the sampling time to the network's input, but this is not guaranteed to capture sampling irregularities during training. Alternatively, the Phased LSTM unit has been created to address this problem by updating its state using the sampling times explicitly.
In this work, we study the effectiveness of the LSTM and Phased LSTM based architectures for the classification of astronomical lightcurves. We use seven catalogs containing periodic and nonperiodic astronomical objects. Our findings show that LSTM outperformed PLSTM on 6/7 datasets. However, the combination of both units enhances the results in all datasets. 
\end{abstract}




\section{Introduction}
Every night, terabytes of photometric data are being collected by wide-field telescopes \citep{bellm2018zwicky}. The amount of data will increase considerably by 2023 when the Vera C. Rubin Observatory will begin operations \citep{ivezic2019lsst}.
These time-domain surveys scan a large portion of the sky, capturing several time-variable phenomena. The data from these variable sources are usually represented as lightcurves that describe underlying astrophysical properties, allowing scientists to identify and label different astronomical objects (\citealt{bonanos2006eclipsing}; \citealt{tammann2008expansion}; \citealt{pietrzynski2018twenty}). For some astrophysical events, it is crucial to obtain this classification in real-time to promptly follow-up individual targets (\citealt{abbott2017multi}; \citealt{abbott2017gravitational}; \citealt{schulze2018cosmic}).
In this context, human visual inspection is not enough to rapidly label incoming data. During the last decade, many Machine Learning (ML) models have been proposed to speed up the automatic classification of objects.
\\\\
Classical ML models have been successfully applied in Astronomy for the classification of photometric data (\citealt{lochner2016photometric}; \citealt{mackenzie2016clustering}; \citealt{lochner2016photometric}; \citealt{castro2017uncertain}; \citealt{devine2018scalable}; \citealt{bai2018machine}; \citealt{martinez2018high}; \citealt{castro2018uncertain}; \citealt{fotopoulou2018cpz}; \citealt{valenzuela2018unsupervised}; \citealt{mahabal2019machine}; \citealt{sanchez2019machine}; \citealt{villar2019supernova}; \citealt{boone2019avocado}; \citealt{zorich2020streaming}; \citealt{sanchez2021alert}). These ``classical" methods require an initial feature extraction module, in which the scientist defines ad hoc features. This procedure is time-consuming and, depending on the features chosen, it could scale the number of operations significantly, implying a high computational cost. Moreover, the features, obtained by expert knowledge, may be insufficient for the models to learn in certain domains e.g., classifying transients using periodic-based features.
\\\\
Since 2015, Deep learning (DL) algorithms have been applied to Astronomy and have outperformed many classical ML models (e.g. \cite{dieleman2015rotation}, \cite{gravet2015catalog}; \cite{cabrera2017deep}; \cite{mahabal2017deep}, \cite{george2018deep}). DL methods aim to automatically extract the features needed to perform a particular task, avoiding the feature engineering process \citep{lecun2015deep}.
\\\\
Recently, recurrent neural networks (RNN, \citealt{rumelhart1986learning}) have been applied to classify sequential data such as lightcurves (\citealt{charnock2017deep}, \citealt{moss2018improved}; \citealt{naul2018recurrent}; \citealt{muthukrishna2019rapid}; \citealt{becker2020scalable}; \citealt{chaini2020astronomical}; \citealt{neira2020mantra}; \citealt{jamal2020neural}) and sequences of astronomical images (\citealt{carrasco2019deep}, \citealt{gomez2020classifying}, \citealt{moller2020supernnova}). However, most DL models are not ideal for processing lightcurves as they assume regular and discrete sampling, while lightcurves measurements have different cadences in a continuous domain. 
\\\\
In the previous paragraph, most of the works use sampling times within the network's input to learn time-related features, which may help with the irregularity. However, even when using time as an input, the weight updates remain regular and discrete, regardless of the spacing between observations. Other approaches try to regularize lightcurve times so we can fit the model assumptions. For example, \citealt{naul2018recurrent} folds lightcurves to decrease the sampling gaps from the continuous timescale, and \citealt{muthukrishna2019rapid} interpolates observations at 3-day intervals. However, folding lightcurves do not assure eliminating the irregularity because it is still an approximation that depends on the number of measurements. On the other hand, interpolation may break variability patterns that come from the astrophysical properties.
\\\\
Alternatively, \citealt{neil2016phased} introduced a novel recurrent unit, called the Phased Long Short Term Memory (PLSTM), an extension of the Long Short Term Memory (LSTM) that incorporates a time gate in charge of considering the effect of irregular sampling on the neural weights updates. We bet on the use of this unit to cover the information lost by the others methods.
\\\\
In this work, we explore the effectiveness of classifying astronomical time series using both the PLSTM and traditional methods, such as the Long Short Term Memory (LSTM), and a Random Forest trained on features. We test our models using OGLE (\citealt{udalski1997optical}), MACHO (\citealt{1995macho}), ASAS (\citealt{pojmanski2005all}), LINEAR (\citealt{stokes2000lincoln}), Catalina Sky Survey (\citealt{drake2009first}), Gaia (\citealt{gaia2016gaia}), and WISE (\citealt{2010AJ....140.1868W}), which are widely-used catalogs containing periodic and non-periodic variable stars. We do not fold the lightcurves for periodical classes because it requires period calculation, which is time-consuming and limits nonperiodical objects' classification.
\\\\
We propose the L+P architecture, a new classifier that combines a PLSTM with an LSTM, the later includes sampling times in its input. We compare this model with a Random Forest (RF) trained on features and the state-of-the-art of the mentioned catalogs. Since our model uses unprocessed observations, it is faster to predict and more accurate than the RF when we have lightcurves with less than 20 observations.
\\\\
This is paper is organized as follows: In Section \ref{sec:background} we provide an overview of the theoretical background of recurrent neural networks, specifically within the context of describing both LSTM and Phased LSTM units. In Section \ref{sec:aboutdata}, we present the data catalogs used in this work. Section \ref{sec:methodology} introduces data preprocessing, the architecture of the proposed model, metrics, and model selection. Finally, we show the results and draw the conclusions in sections \ref{sec:results} and \ref{sec:conclusions}, respectively.

\section{Background}\label{sec:background}
%
%
Neural networks (NN) are parametric functions capable of transforming a vector input $\boldsymbol{x} \in \mathbb{R}^D$ to an expected target $\boldsymbol{y}$. We will focus on a classification task, where the output consists of one of $K$ classes. Hereafter, we assume $\boldsymbol{y} \in \mathbb{R}^K$ is a one-hot encoded vector that is 1 for the actual class and zero otherwise.
\\\\
A NN comprises neurons that transform the input by multiplying it by a weight matrix $\rm{W}$ and adding a vector of biases $\boldsymbol{b}$. This transformation is passed through an element-wise nonlinear activation function $g$. When training a NN, the goal is to learn $\rm{W}$ and $\boldsymbol{b}$, capturing nonlinear relationships between variables. Equation \ref{eq:fc} describes a Fully-Connected (FC) layer which receives $\boldsymbol{x}$ as input,
\begin{align}
 & \boldsymbol{h} = g(\mathrm{W}^{\top} \boldsymbol{x} + \boldsymbol{b}),\label{eq:fc}
\end{align}
where $\boldsymbol{h}$ is the layer output. Stacking FC layers increases the classification capacity of the NN by learning features at multiple levels of abstraction \citep{glorot2010understanding}. In this case, the output of each stacked layer serves as input for the next one. These particular types of neural networks are called multi-layer perceptron (MLP) \citep{rumelhart1986learning}.
\\\\
MLPs have been widely used for regression and classification tasks. Their matrix representation allows fast computation, which can be parallelized on GPU \citep{oh2004gpu}. Usually, MLPs assume that each entry within the input is independent of each other. However, this is not the case of a time series, where the input is a sequence of measurements, and each point may be conditioned on the previous points.
\\\\
In the following sections we explain the architecture of the recurrent neural network (RNN), an extension of the MLPs capable of learning time-based patterns from data (Section \ref{back:rnn}). We also describe the operation of the LSTM (Section \ref{back:lstm}) and PLSTM (Section \ref{back:plstm}), as they are the two types of recurrent neurons used in this work.

\subsection{Recurrent Neural Networks}\label{back:rnn}
\begin{figure*}
    \centering
    \includegraphics[scale=0.36]{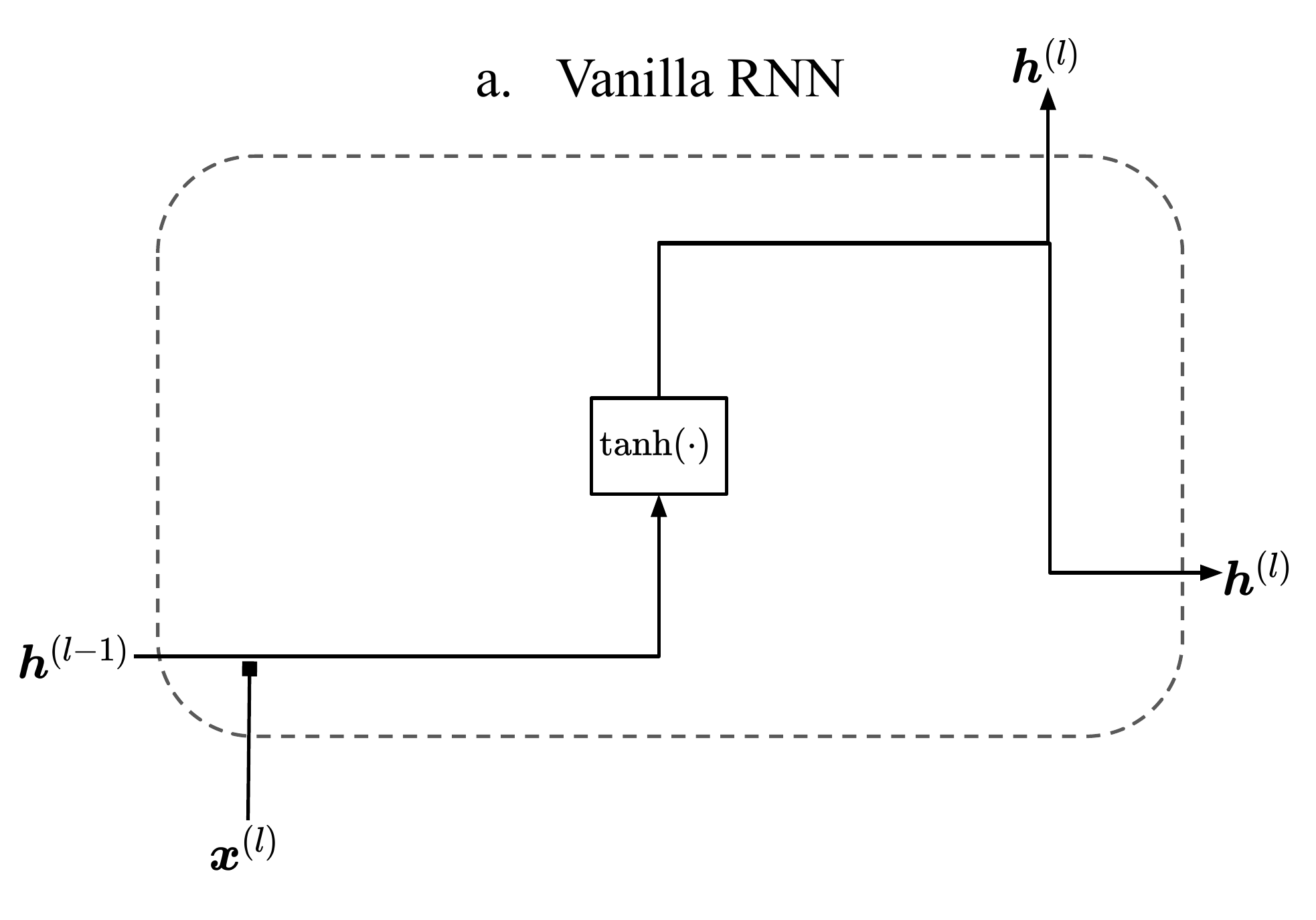}
    \includegraphics[scale=0.36]{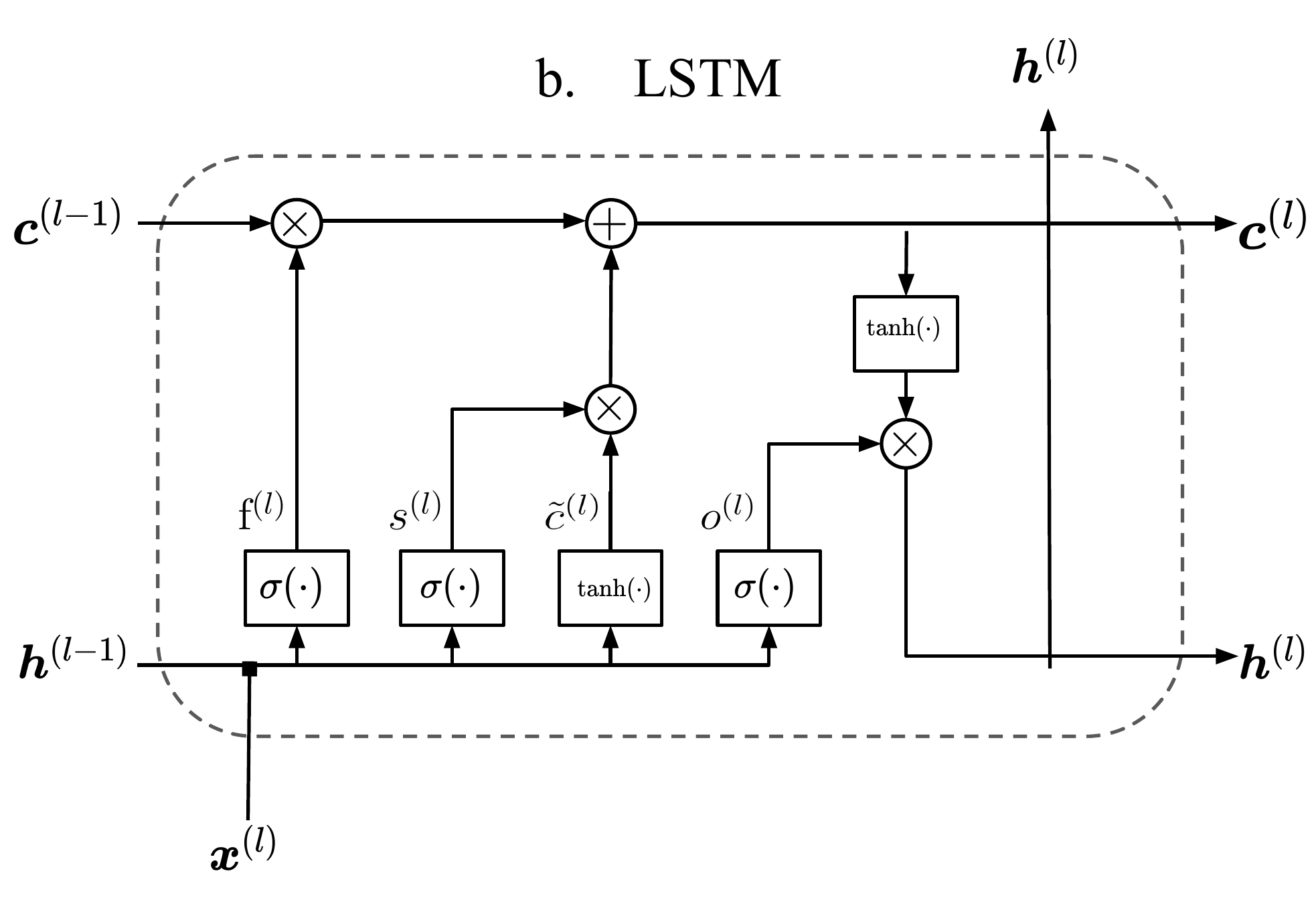}
    \includegraphics[scale=0.36]{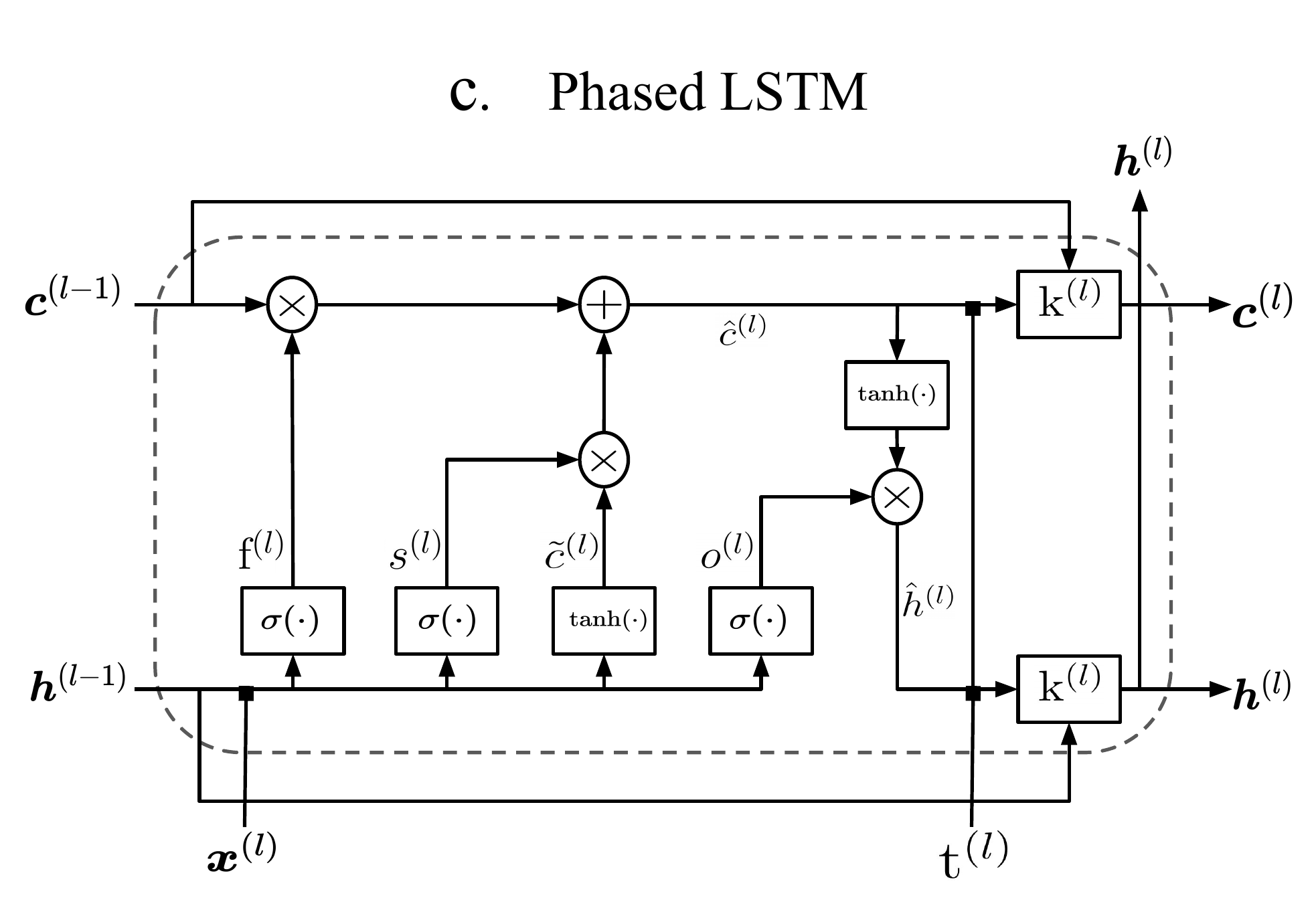}
    \caption{Recurrent neural units. Everything within the dashed line is part of the neuron. The rows represent the flow of information, starting from the inputs ($\boldsymbol{h}^{(l-1)}$ previous hidden and $\boldsymbol{c}^{(l-1)}$ cell state, $\boldsymbol{x}^{(l)}$ current observation and $t^{(l)}$ its sampling time) outside the neuron to the new states ($\boldsymbol{h}^{(l)}$ hidden, and $\boldsymbol{c}^{(l)}$ cell state). Boxes are nonlinear transformation for recurrent gates ($f^{(l)}, s^{(l)}, o^{(l)}, k^{(l)}$) and the candidate state ($c^{(l)}$) with sigmoid ($\sigma(\cdot)$) and hyperbolic tangent ($\tanh(\cdot)$) activation functions respectively.}
    \label{fig:recurrent_units}
\end{figure*}

%
%
Recurrent neural networks are capable of learning dependencies on sequential data. They use a hidden state vector $\boldsymbol{h}$ that preserves the information of previous time steps. Formally, a vanilla recurrent unit is given by:
\begin{align}
& \boldsymbol{h}^{(l)} = g{\left(\mathrm{W}^{\top}_x \boldsymbol{x}^{(l)} + \mathrm{W}_h^{\top} \boldsymbol{h}^{(l-1)} +\boldsymbol{b} \right)} , \label{eq:vanilla_unit}
\end{align}
where $[\mathrm{W}^{\top}_x$, $\mathrm{W}^{\top}_h]$ are the weights for $x$ the input and $h$ the hidden state at the current ($l$) and previous ($l-1$) time step, respectively. Note the representation is quite similar to an MLP. However, recurrent neurons combine  the current observations $\boldsymbol{x}^{(l)}$ with  the previous activation (or hidden state) $\boldsymbol{h}^{(l-1)}$ to build the input to the $l$-th layer.
\\\\
Figure \ref{fig:recurrent_units}a shows a vanilla recurrent neural unit corresponding to Equation \ref{eq:vanilla_unit}. Notice that the information flows in the sense of arrows. 
The arrow bifurcation in the last part of the diagram indicates the potential of the neuron to make predictions while connecting with the next step. In other words, the hidden state $\boldsymbol{h}^{(l)}$ represents a feature vector that describes the time series up to the current step. Usually, RNNs use a FC layer that receives the hidden state to classify new observations,
\begin{align}
    \boldsymbol{\hat{y}^{(l)}} = \text{softmax}(\rm{W_o^{\top}}\boldsymbol{h^{(l)}} + \boldsymbol{b}_o)
\end{align}
where $\rm{W}^{\top}_o$ and $\boldsymbol{b}_o$ are the weights and biases of the FC layer, and $\boldsymbol{\hat{y}^{(l)}}$ is the predicted label at the $l$-th step.
\\\\
During the training of RNNs, the gradients flow through the steps of the network. To calculate the gradients, we differentiate the loss function regarding the activation of each time step. Thus, if the gradient is backpropagated until the first time step, as the length of the sequence increases, several matrix multiplications are performed because of the chain rule \citep{werbos1990backpropagation}. If at some time in the recurrence, the gradients have low values ($< 1$), the total product may vanish. On the other hand, if the values are high ($> 1$), the gradient could explode. This problem is known as \textit{the exploding and vanishing of the gradient} \citep{pascanu2012understanding} and is the motivation for creating the Long Short Term Memory (LSTM) recurrent unit. 
%
%
\subsection{Long Short Term Memory (LSTM)}\label{back:lstm}
The Long Short Term Memory unit, introduced by \citealt{hochreiter1997long}, is a recurrent unit capable of learning long-term dependencies from time series. Unlike the vanilla recurrent unit, the LSTM has an additional cell state that scales the neuron's output according to long-term relevant information stored in it. 
\\\\
Both the LSTM cell ($\boldsymbol{c}^{(l)}$) and the hidden ($\boldsymbol{h}^{(l)}$) state vectors have the same dimensionality, but their values are updated differently. Updates are controlled by recurrent gates, MLPs that receive the current observation and the previous hidden state to learn specific tasks. Typically, the LSTM uses a gate to save ($\boldsymbol{s}^{(l)}$) and another one to forget ($\boldsymbol{f}^{(l)}$) long-term patterns.  A third gate ($\boldsymbol{o}^{(l)}$) scales the output of the unit. Formally, 
\begin{align}
    & \boldsymbol{c}^{(l)} = \boldsymbol{f}^{(l)} \odot \boldsymbol{c}^{(l-1)} + \boldsymbol{s}^{(l)} \odot \boldsymbol{\tilde{c}} \label{cellstate_update}\\
    & \boldsymbol{h}^{(l)} = \boldsymbol{o}^{(l)} \odot \tanh{(\boldsymbol{c}^{(l)})} \label{hiddenstate_update}
\end{align}
where $\odot$ is the Hadamard product of vectors, $\boldsymbol{\tilde{c}}$ is a candidate cell state defined as,
\begin{align}
 & \boldsymbol{\tilde{c}}^{(l)} = \tanh{(\rm{W}_{cx}^{\top}\boldsymbol{x}^{(l)} + \rm{W}_{ch}^{\top}\boldsymbol{h}^{(l-1)} + \boldsymbol{b_c})}
\end{align}
 and
\begin{align}
			& \boldsymbol{f}^{(l)} = \sigma(\rm{W}_{fx}^{\top}\boldsymbol{x}^{(l)}+\rm{W}_{fh}^{\top}\boldsymbol{h}^{(l-1)}+\boldsymbol{b_f}) \label{eq:forget}\\
			& \boldsymbol{s}^{(l)} = \sigma(\rm{W}_{sx}^{\top}\boldsymbol{x}^{(l)}+\rm{W}_{sh}^{\top}\boldsymbol{h}^{(l-1)}+\boldsymbol{b_s}) \label{eq:save}\\
			& \boldsymbol{o}^{(l)} = \sigma(\rm{W}_{ox}^{\top}\boldsymbol{x}^{(l)}+\rm{W}_{oh}^{\top}\boldsymbol{h}^{(l-1)}+\boldsymbol{b_o}). \label{eq:output}
\end{align}
In this formulation, $[\rm{W}_{fx}, \rm{W}_{fh}, \boldsymbol{b_f}]$, $[\rm{W}_{sx}, \rm{W}_{sh}, \boldsymbol{b_s}]$, and $[\rm{W}_{ox}, \rm{W}_{oh}, \boldsymbol{b_o}]$ are the weights and biases of the forget, save, and output gate, respectively.
\\\\
Figure \ref{fig:recurrent_units}b. shows the LSTM structure and its operations. The network's input is formed by the new observations and the previous hidden and cell states, typically initialized to zero.
\\\\
LSTM signified a breakthrough for long time series analysis in many fields of science (\citealt{hochreiter1997lstm}; \citealt{greff2016lstm}). However, this unit does not consider irregular sampling times when updating the states in equations \ref{cellstate_update} and \ref{hiddenstate_update}. Consequently, it assumes that all step updates weigh the same regardless of how far they are from the last observation.
%
%
\subsection{Phased Long Short Term Memory (PLSTM)}\label{back:plstm}
\begin{figure}
    \centering
    \includegraphics[scale=0.6]{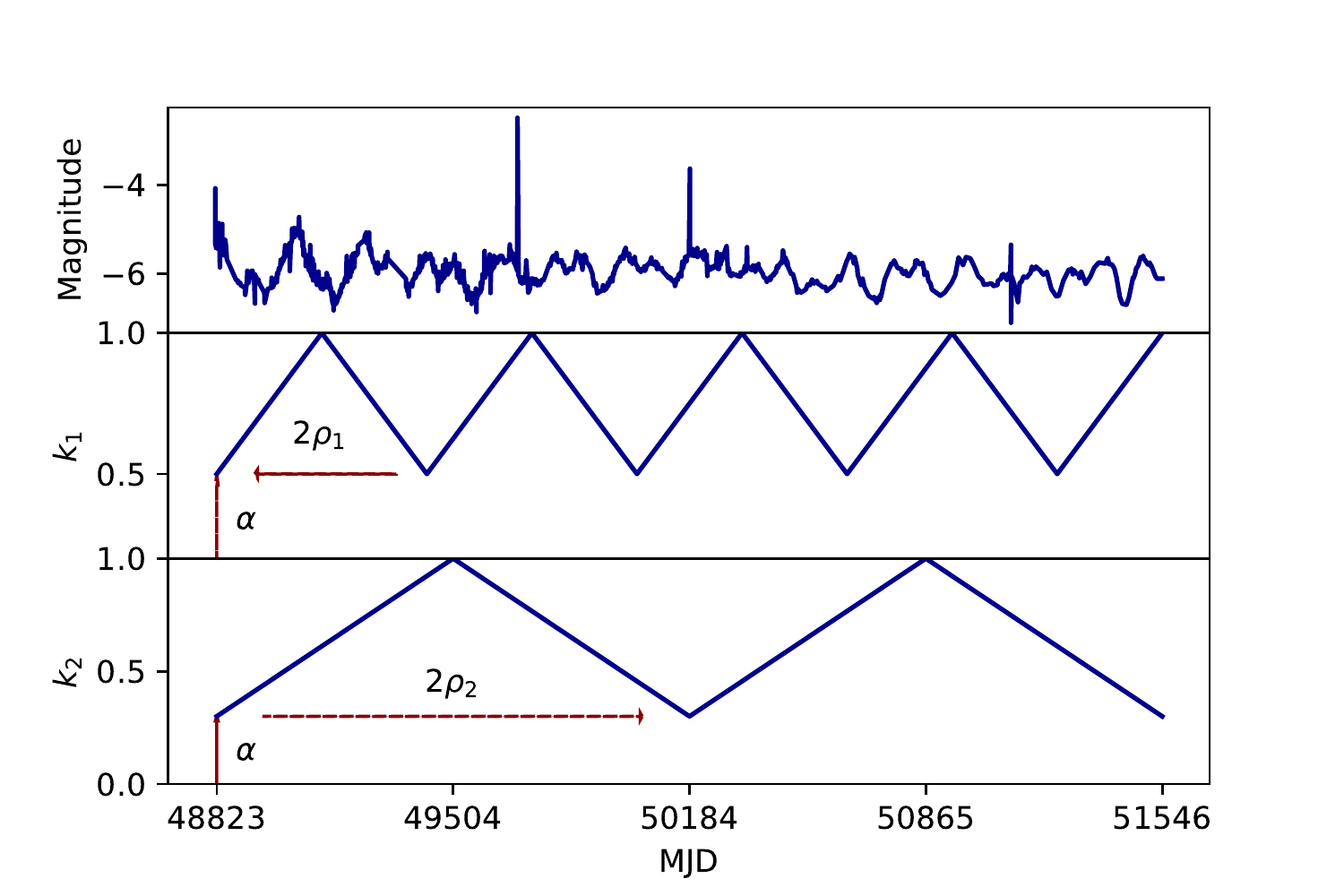}
    \caption{Time gate openness phases controlling the update rate for the states of a single lightcurve. The trainable openness duration $\rho$ depends on a particular neuron, while the leak parameter $\alpha$ is shared across the state. A neuron is fully updated when the curve is in the peak, i.e., $k_{i} = 1$. Otherwise, it is partially updated, as described in Equation  \ref{eq:timegate}.}
    \label{fig:timegatephase}
\end{figure}

The Phased Long Short Term Memory proposed by \citealt{neil2016phased} is an extension of the LSTM unit for processing inputs sampled at irregular times. The idea is to consider sampling times explicitly during the state updates. To do this, the PLSTM assumes the observations come from periodic sampling and, therefore, the neurons should be controlled by independent rhythmic oscillations following these periodicities.
\\\\
Every neuron in the H-dimensional state has a learnable independent period, forming the vector $\boldsymbol{\tau}  \in \mathbb{R}^H$. Then, for a given $l$-th step, we calculate  the phase of neurons $\boldsymbol{\phi}^{(l)}$ as
\begin{align}\label{eq:openness_phase}
    & \boldsymbol{\phi}^{(l)} = \frac{(t^{(l)} - \boldsymbol{s})\ mod\ \boldsymbol{\tau}}{\boldsymbol{\tau}},
\end{align}
where $\boldsymbol{s} \in \mathbb{R}^H$ represents the trainable shift of the signal and $t^{(l)}$ is the $l$-th sampling time. PLSTM adds a time gate which is defined as a piece-wise function depending on the phases and a trainable vector $\boldsymbol{\rho} \in \mathbb{R}^{H}$ that controls the duration of the openness phase,
\begin{align}\label{eq:timegate}
    & \boldsymbol{k}^{(l)} = \begin{cases} 
      \frac{2\boldsymbol{\phi}^{(l)}}{\boldsymbol{\rho}} & if\ \boldsymbol{\phi}^{(l)}< \frac{1}{2}\boldsymbol{\rho} \\
      2-\frac{2\boldsymbol{\phi}^{(l)}}{\boldsymbol{\rho}} & if \frac{1}{2}\boldsymbol{\rho} < \boldsymbol{\phi}^{(l)} < \boldsymbol{\rho} \\
      \alpha\boldsymbol{\phi}^{(l)} & otherwise .
   \end{cases}
\end{align}
In Equation \ref{eq:timegate} $\alpha$ is a leak parameter close to zero which allows gradient information flow \citep{he2015delving} for those cases where the gate is closed. The time gate $k^{(l)}$ controls how much of the proposed states should flow to the next recurrence. Unlike the LSTM, the PLSTM  does not consider every update equally important to each other, and therefore, it weighs the updates by using the current sampling time. Using the output of the time gate formulation in Equation \ref{eq:timegate}, we update the cell states in the following way: 
\begin{align}
    & \hat{\boldsymbol{c}}^{(l)}  = \boldsymbol{f}^{(l)} \odot \boldsymbol{c}^{(l-1)} + \boldsymbol{s}^{(l)} \odot \boldsymbol{\tilde{c}}^{(l)} \\
  & \label{eq:cellstate} \boldsymbol{c}^{(l)}   = \boldsymbol{k}^{(l)} \cdot \boldsymbol{\hat{c}}^{(l)} + (1-\boldsymbol{k}^{(l)}) \cdot \boldsymbol{c}^{(l-1)} \\
   &  \hat{\boldsymbol{h}}^{(l)}  = \boldsymbol{o}^{(l)} \odot g(\boldsymbol{c}^{(l)}) \\
  & \label{eq:hiddenstate} \boldsymbol{h}^{(l)}   = \boldsymbol{k}^{(l)} \cdot \boldsymbol{\hat{h}}^{(l)} + (1-\boldsymbol{k}^{(l)}) \cdot \boldsymbol{h}^{(l-1)}
\end{align}
where $\boldsymbol{\hat{c}}_l$ and $\boldsymbol{\hat{h}}_l$ are the candidate states at $l$-th step. Note that if we replace $\boldsymbol{k}^{(l)} = 1$, then we get the LSTM output.
\\\\
For irregular sampled continuous-time sequences, the observation time controls the update of each neuron (see Figure \ref{fig:timegatephase}), allowing improvements in the convergence ratio as well as the loss minimization.
The Phased LSTM has also been used for multi-event time series in which independent sequences describe the same phenomena (\citealt{anumula2018feature}; \citealt{liu2018learning}). 

\section{Data}\label{sec:aboutdata}
Our classification domain is composed of variable stars in the form of photometric lightcurves. Each light curve corresponds to a set of observations taken at different times and varies in length depending on the objective of the surveys. In this context, we collected data from seven widely used surveys that we describe as follows:
\\\\
\noindent\underline{The Massive Astrophysical Compact Halo Objects (MACHO):}\vspace{2mm}
The MACHO project \citep{alcock2000macho} results from a collaboration between different institutions to look for dark matter in the halo of the Milky Way. Based on several years of observation, astronomers tried to find the dark matter by its effect on the bright matter. We used the catalog from \citealt{kim2011quasi}. Table \ref{tab:macho_dataset} shows the MACHO dataset composition following the same blue-band selection made by \citealt{nun2014supervised}. The objects are principally periodic variable stars except for Quasars (QSO) and Microlensings (MOA).
\\\\
\noindent\underline{The Lincoln Near-Earth Asteroid Research (LINEAR):}\vspace{2mm}\\
LINEAR \citep{stokes2000lincoln} is a collaborative program aiming at monitoring asteroids, founded by the US Air Force and NASA, and has the collaboration of The Lincoln Laboratory at the Massachusetts Institute of Technology. Table \ref{tab:linear_dataset} shows the training set composition made by \citealt{naul2018recurrent}, which contains periodical lightcurves measured in $r\leq18$ red band (see \citealt{palaversa2013exploring} for more details).
\\\\
\noindent\underline{The All Sky Automated Survey (ASAS):}\vspace{2mm}\\
With two observation stations, ASAS \citep{pojmanski2005all} is a low-cost project focused on photometric monitoring of the entire sky. A catalog of variable stars constitutes one of the main goals of this project. We use V-band periodical lighcurves selected by \citealt{naul2018recurrent}. Table \ref{tab:asas_dataset} describes the number of object per class.
\\\\
\noindent\underline{The Optical Gravitational Lensing Experiment (OGLE):}\vspace{2mm}\\
OGLE \citep{udalski1997optical} is an observation project, currently operational (OGLE-IV), that started in 1992 (OGLE-I) under the initiative of the University of Warsaw. Like MACHO, this project studies the dark matter in the Universe using microlensing phenomena. In 2001, the third phase (OGLE-III) of the project began by including observations at Las Campanas Observatory, Chile \citep{udalski2004optical}. We used a selection made by \citealt{becker2020scalable} shown in Table \ref{tab:ogle_dataset} corresponding to I-band lightcurves.
\\\\
\underline{Wide-field Infrared Survey Explorer (WISE):}\vspace{2mm}\\
The WISE project \citep{2010AJ....140.1868W} collects observations of the whole sky in four photometric bands. Approximately six months was necessary to complete the first full coverage. Since then, the telescope was intermittently used mainly to detect asteroids. In 2013, a release including the whole dataset was published with their corresponding catalog and other tabular data. In this work, we used the W1-band selection (\citealt{nikutta2014meaning}) of lightcurves used by \citealt{becker2020scalable}. The dataset composition is shown in Table \ref{tab:wise_dataset}.
\\\\
\underline{Gaia DR2 Catalog of variable stars (Gaia):}\vspace{2mm}\\
The Gaia mission \citep{gaia2016gaia}, promoted by the European Space Agency, is a collaboration to chart a three-dimensional map of the galaxy to study the structure, composition, and evolution of the Milky Way. The second data release (DR2) \citep{gaia2018gaia} includes observations for more than 1.3 billion objects in three photometric bands of which we only use the G-passband. As in WISE and OGLE, we used the Becker's selection \citep{becker2020scalable} of lightcurves. The objects used in this work are described in Table \ref{tab:gaia_dataset}.\\\\
\\\\
\underline{The Catalina Sky Survey (CSS):}\vspace{2mm}\\
CSS \citep{drake2009first} is a project led by NASA and is fully dedicated to discovering and tracking near-Earth objects.  Specifically, we worked with the second release, which contains photometries derived from seven years of observation. We combined south and northern variable stars with transient objects, being the most heterogeneous dataset of this work. All the objects were measured in the V-passband. Due to the diversity of objects, the cadences deviation is higher than in other surveys, as shown in Table \ref{tab:cadences}, and the number of samples per class is hugely imbalanced. We balanced the training set by undersampling the most numerous classes up to 300 and removing those with less than 100 samples, -i.e., Cepheid Type I (CEPI), Low-amplitude $\delta$ Scutis (LADS), Post-common-envelope binaries (PCEB), and Periodic variable stars with unclear behavior (Hump). The final dataset composition can be seen in Table \ref{tab:cata_dataset}.
\input{Tables/data_tables}
\begin{table}
    \centering
    \begin{tabular}{c|c|c}
        Dataset & Cadence [days]\\ \hline
        MACHO   & 2.65  $\pm$ 6.64\\
        OGLE    & 3.70  $\pm$ 12.88\\
        WISE    & 4.01  $\pm$ 25.53\\
        ASAS    & 5.59  $\pm$ 19.76\\
        LINEAR  & 8.41  $\pm$ 31.41\\
        CSS     & 14.67 $\pm$ 42.26\\
        Gaia    & 26.82 $\pm$ 32.33\\
         \hline
    \end{tabular}
    \caption{Survey cadences in days}
    \label{tab:cadences}
\end{table}

\section{Methodology}\label{sec:methodology}
In this section, we describe our models, architectures and training methodology used in this work. First, we introduce the preprocessing step to prepare the data as input for the recurrent model (Section \ref{subsect:preprocessing}). We describe two normalization techniques, the zero-padding strategy for dealing with the variable-length problem, and a resampling routine to decrease the number of computations within the network loop. Then, we present the training architecture and the combination of the recurrent units to predict for new unobserved lightcurves (\ref{subsec:arch}).
Finally, we explain the metrics used (Section \ref{subsec:metrics}) and the model selection strategy (Section \ref{modelselection}), which are invariant to the number of samples per class, addressing the consequences of training with imbalanced datasets.
\subsection{Data Preprocessing}\label{subsect:preprocessing}
Lightcurves in the form of time series fed the input of the network. Each lightcurve is a matrix of $L \times D$ observations, where $L$ is the length or number of time steps, and $D$ represents the dimensionality of the input parameters. In this work, we used observation times, the mean, and the uncertainty of the magnitudes, hence $D=3$.
\\\\
As a preprocessing step, we start by scaling and shifting the input parameters to help the optimization process. (\citealt{shanker1996effect}, \citealt{jayalakshmi2011statistical}). Many studies in the literature use folded lightcurves where the observations are shifted to the phase domain, facilitating the recognition of patterns. However, the folding process is time-consuming because it requires knowledge of the period, which is $O(L\log(L))$ (\citealt{vanderplas2018understanding}) and $O(L)$ approximated (\citealt{mondrik2015multiband}). Moreover, this only works on periodic signals, and we are also interested in classifying nonperiodic objects. Thus, the time complexity of the proposed method should be insignificant concerning the forward pass of the network. It should also be as general as possible, avoiding any assumptions on the domain, such as periodicity. Without these considerations, we would be falling into the same problems of models trained on features.
\subsubsection{Normalization}\label{sec:normalizations}
\begin{figure*}
    \centering
    \includegraphics[scale=0.25]{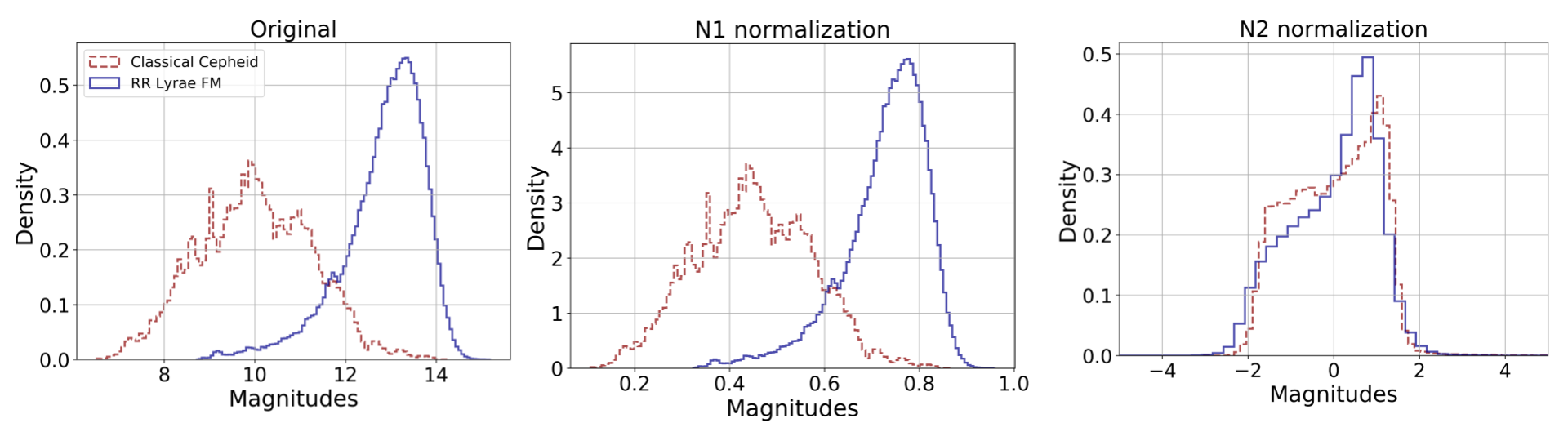}
    \caption{Cepheids and RR Lyrae magnitude distributions using different normalization methods. \textbf{N1} normalizes using the minimum and maximum values across all lightcurves. \textbf{N2} normalizes each sample separately, so they get zero mean and unit standard deviation.}
    \label{fig:n1vsn2}
\end{figure*}

We define two methods to scale and shift the observations. The first normalization method (N1) uses a min-max scaler along the samples. The second normalization (N2) calculates the z-score of each sample separately to have zero mean and unit standard deviation. The best normalization method for a given dataset is selected after training as a hyperparameter of the model. The effect of the normalization procedure may have a strong impact on the results of our models. As an example, Figure \ref{fig:n1vsn2} shows the distribution of the lightcurve magnitudes of Cepheids and RR Lyrae, from the ASAS dataset. The N1 normalization keeps the bimodality of the classes, which is more straightforward than N2, where the network is forced to separate by time dependencies instead of the current magnitude.
\subsubsection{Padding Lightcurves}
\begin{figure}
    \centering
    \includegraphics[scale=0.35]{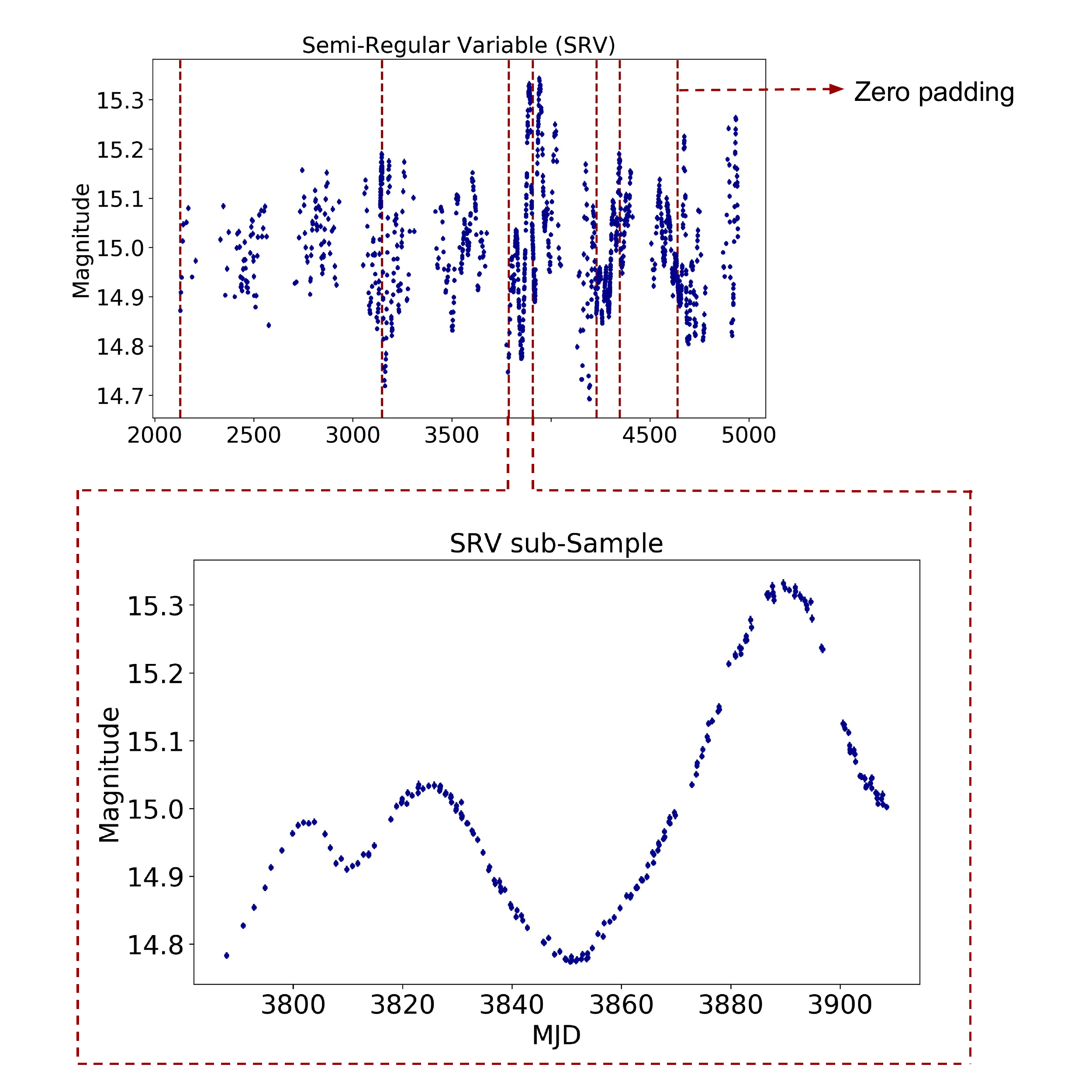}
    \caption{Masking and split preprocess step. A semi regular variable (SRV) from OGLE-III dataset is shown on the top chart. On the bottom figure, a zoom over a subsample of 200 observations is displayed in order to visualize the new input of the network. If we have less than 200 observations, we fill the last split with zeros (see the zero-padding portion in the top figure).}
    \label{fig:padding}
\end{figure}

Working with real lightcurves implies dealing with the variable-length problem of time series. The number of observations between astronomical objects differs and, therefore, it is not possible to train a matrix-based model such as neural networks. Consequently, we need to standardize the length among samples.
\\\\
The most typical solutions rely on the interpolation and padding of lightcurves. Interpolation consists of estimating intermediate values in the observation sequence. On the other hand, the padding technique equalizes the lengths by inserting filler values (typically zero) and removing them during loss calculation. This technique is also known as zero-padding. In this work, we use the padding technique since interpolation could affect the underlying astrophysics properties of the lightcurves.
\\\\
Even though zero padding solved the variable-length problem, a large variance in the number of observations could be inefficient in memory management. For example, on the ASAS dataset, the longest lightcurve has 1745 observations, while the shortest has 7 observations (see Table \ref{tab:asas_dataset}).  Using the padding technique implies masking 1738 dummy observations for the shortest lightcurve or, in other words, 1738 loop-operations without contribution to training. 
\\\\
We sampled from the original lightcurves to reduce the variance in the number of observations, as shown in Figure \ref{fig:padding}. Each dashed line on the top chart defines a subset. The input of the network will be $L'$ consecutive time steps such that $L'<L$ the original number of observations. In this work, we defined $L'=200$. It is important to highlight we only sample during training, so we use the full length lightcurves on testing. The subsampling process is also an alternative to the truncated backpropagation \citep{werbos1990backpropagation}, which decomposes the gradient as a sum of the losses at each time step \citep{puskorius1994truncated}.
\\\\
\subsection{Neural Network Architecture}\label{subsec:arch}
\begin{figure}
    \centering
    \includegraphics[scale=0.42]{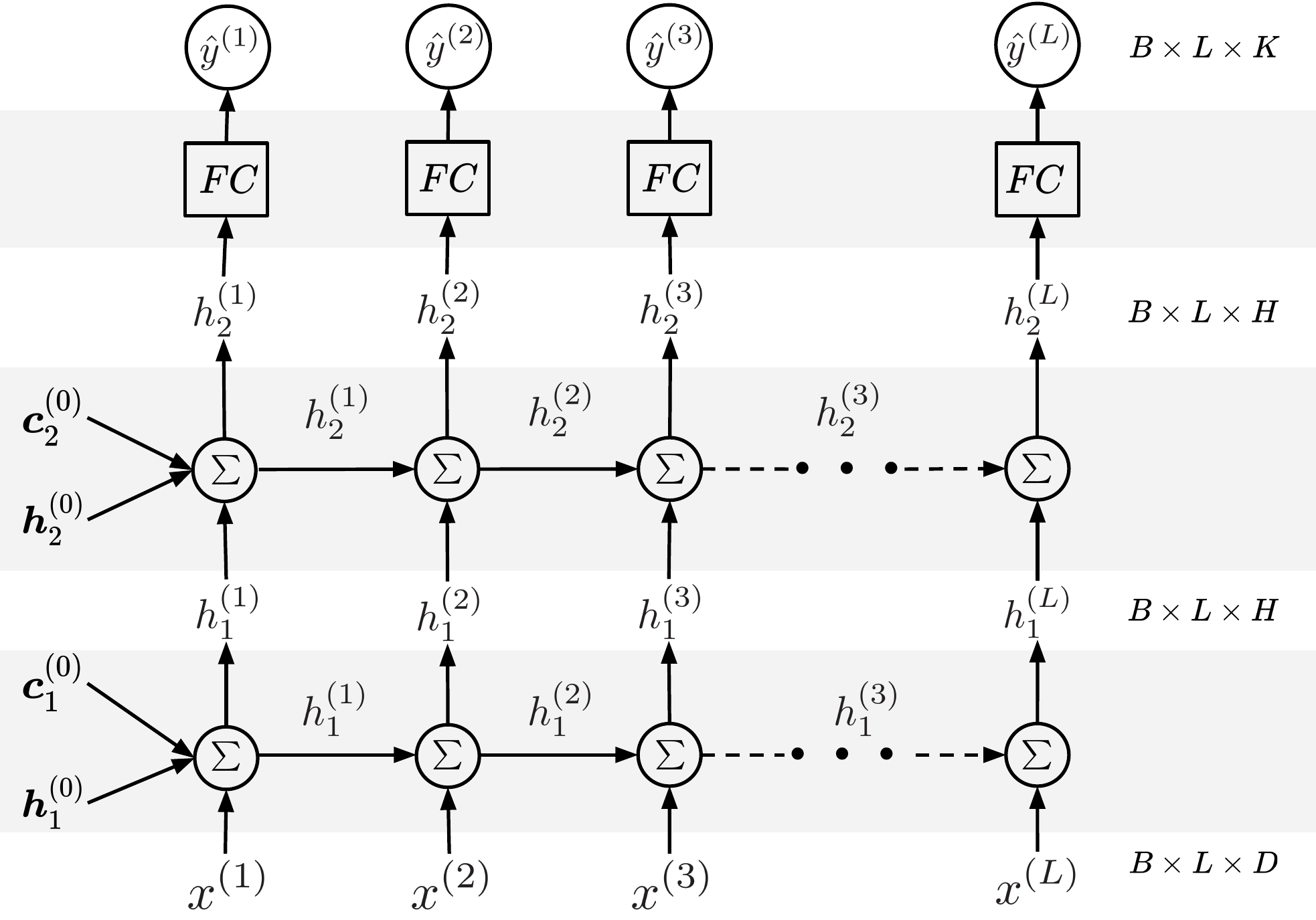}
    \caption{Recurrent network architecture. $B$, $L$, $H$, $D$ and $K$ are the batch size (or number of samples), time steps, hidden units, input parameters, and number of classes, respectively. The $\sum$ symbol represents the recurrent unit operations. The first layer receives the input matrix $x^{(l)}$ with $l = \{0, ..., L\}$ and the states $c^{(l)}$, $h^{(l)}$. Note that the 0 superscript at the beginning defines the initial states while the recurrent layer is defined by the subscript. In the third layer we used a fully connected layer ($FC$) + softmax to obtain the class probabilities. }
    \label{fig:netarch}
\end{figure}

\begin{figure}
    \centering
    \includegraphics[scale=0.4]{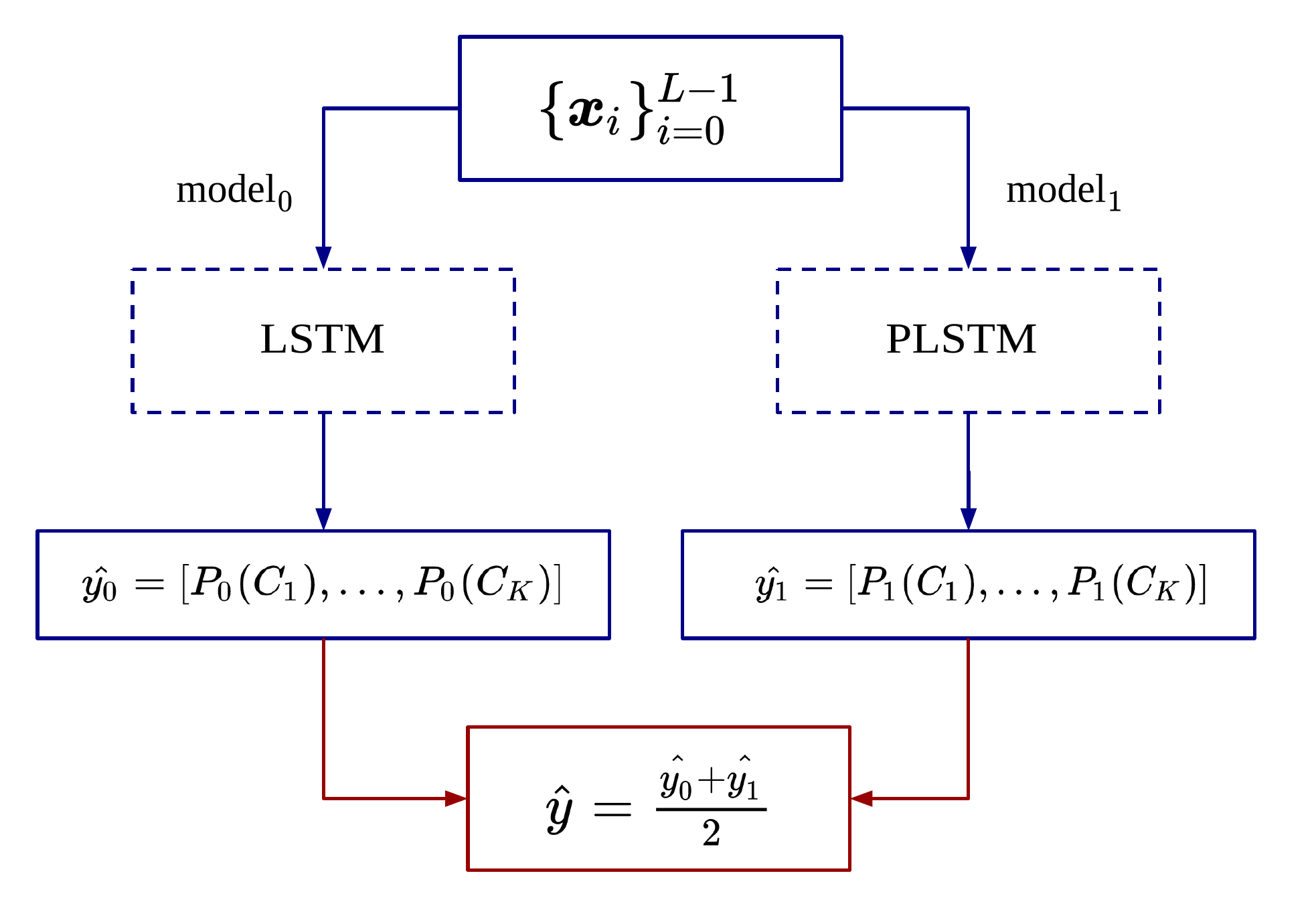}
    \caption{Testing architecture based on the combination of LSTM and PLSTM adjusted models.}
    \label{fig:ensemble}
\end{figure}

The network architecture is presented in Figure \ref{fig:netarch}. We use two unidirectional layers of RNN, where $\sum$, the recurrent unit, is either LSTM or PLSTM. Both layers have the same number of neurons ($H = 256$) and are initialized with zero states $(c_{0}, h_{0})$. We use a batch size of B=400 samples for training. Notice that an epoch consists of $N/B$ iterations, where $N$ is the number of samples.
\\\\
We apply layer normalization according to \citealt{ba2016layer} over each recurrent unit output. Additionally, we use a dropout \citep{semeniuta2016recurrent} of 0.5 probability over the second-layer output. The last part of the architecture is a Fully Connected ($FC$) classifier, which uses the hidden state to build a $K$-dim probability vector;  $K$ is the number of classes.   
\\\\
After training, we fix the best weight configuration for the LSTM and PLSTM classifiers, and we compute the forward pass to generate [$\boldsymbol{\hat{y}_0}$, $\boldsymbol{\hat{y}_1}$], the corresponding probability vectors, as shown in Figure \ref{fig:ensemble}. The final prediction $\boldsymbol{\hat{y}}$ is the average between the predictions. We named this testing setup the L+P architecture.
\subsection{Backpropagation}
The backpropagation step \citep{boden2002guide}, starts with the loss estimation that was carried out by the categorical cross-entropy (CCE) function (also called softmax loss). For each $b$-th batch sample, we calculate the corresponding loss at each $l$-th time step as follows: 
\begin{align}\label{eq:cce}
    & \mathcal{L_{CCE}}(\hat{\rm{Y}}_{B\times L\times K}) = -\frac{1}{B}\sum_{b=1}^B\sum_{l=1}^{L}\sum_{k=1}^K y_{bk}\log(\hat{y}_{blk}).
\end{align}
In the above formulation, $\boldsymbol{y}_b \in \mathbb{Z}^K$ is a one-hot vector associated with the class label of each $b$-th sample, and $\boldsymbol{\hat{y}}_{bl} \in \mathbb{R}^K$ is prediction vector at the $l$-th time step. Using the sum of losses along the $L$ time steps forces the network to predict as soon as possible, getting better classification results when we have fewer observations in the lightcurve.
\\\\
Finally, we applied \textit{Adaptive Moment Estimation} (ADAM) introduced by \citealt{kingma2014adam} with an initial \textit{learning rate} of 0.001 to update the network weights. The learning curves associated with the validation loss are presented in Appendix \ref{appendix:lc}.
\subsection{Evaluation Metrics}\label{subsec:metrics}
Since the training sets are imbalanced, we employed the macro F1 score \citep{van2013macro}, a harmonic mean of precision and recall calculated separately for each $k \in K$ class, 
\begin{align}
    & \textrm{F1}  = \frac{1}{K} \sum_{k=0}^{K-1} 2 \times \frac{\textrm{Precision}_k \times \textrm{Recall}_k}{\textrm{Precision}_k + \textrm{Recall}_k}
    \nonumber
\end{align}
where, 
\begin{align}
    & \textrm{Recall}_k  =   \frac{\textrm{True Positives}_k}{\textrm{True Positives}_k + \textrm{False Negatives}_k}
    \nonumber\\
    & \textrm{Precision}_k  =    \frac{\textrm{True Positives}_k}{\textrm{True Positives}_k + \textrm{False Positives}_k}.
    \nonumber
\end{align}
\\\\\
The precision score determines how many of our predicted labels are actually true, and the recall indicates the number of true labels we were able to identify correctly. Using the macro average, all classes contribute equally regardless of how often they appear in the dataset.
\subsection{Random Forest Classifier}
As mentioned in previous sections, we used a Random Forest (RF) trained on features for comparison purposes. An RF is a classical machine learning method that combines the outputs of several decision trees for predicting \citep{breiman2001random}. Each decision tree splits the feature space to create isolated subgroups that separate objects with similar properties. We usually run a feature extraction process on the observations to characterize sequences using a standard set of descriptors. In this work, we use a python library published by \citep{nun2015fats} called FATS (Feature Analysis for Time Series), which facilitates the extraction of time-series features, such as the period and the autocorrelation function.  Table \ref{tab:fatsfeatures} shows all the single-band features we selected for training. Because of library constraints, we only calculate features on lightcurves with at least ten observations.  
\subsection{Model Selection}\label{modelselection}
We use k-fold cross-validation, with k=3. Each k-fold is composed of training (50\%), validation (25\%), and testing (25\%) random subsets. We use the validation set during training to see how well our recurrent classifier generalizes over unseen objects. We perform the validation step at the end of each training epoch with a maximum of 2000 repetitions. However, we can stop the training process if the validation loss does not decrease compared to the best historical loss. In this case, we set 30 epochs as patience for the early stopping. In the case of RF, the validation set was used to find hyperparameters.
\\\\
After $k$-fold training, we test our models using all available observations. For RNNs, we use the last hidden state of every lightcurve to build the final prediction. The best model is the one that maximizes the mean F1 scores of the cross-validated models.

\section{Results}\label{sec:results}
This section compares our model based on LSTM and PLSTM recurrent units with a RF trained on features and other models from the literature (\citealt{naul2018recurrent}; \citealt{becker2020scalable}; \citealt{zorich2020streaming}). 
\\\\
Evaluation was shown to compute metrics using all available observations in the lightcurve (offline) and incrementally in-time (streaming). We observed significant improvements with the proposed recurrent model to classify lightcurves with less than 20 observations, being an excellent alternative to classify objects with less number of points in their lightcurves.
\\\\
Next, we discuss the benefits of using recurrent models against an RF regarding the feature extraction process and time complexity. Confusion matrices and learning curves are shown in the Appendix \ref{appendix:cm}. 
\subsection{Offline Evaluation}\label{offline}
We start by testing models using the whole lightcurve and evaluating them at the last time step. According to our cross-validation strategy (see Section \ref{modelselection}), we ran each model 3 times. For each architecture, we select the type of normalization that maximizes the mean F1-score. Table \ref{tab:summary} summarizes the performance of each model based on their precision, recall and F1-score.
\\\\
For OGLE, WISE, Gaia, and LINEAR, the F1 score of the single LSTM and PLSTM are higher than that of RF. For MACHO, CSS, and ASAS, the PLSTM performs worse than the RF; however, the LSTM models outperform the RF. Combining PLSTM and LSTM recurrent neural networks works better in all datasets, obtaining the best performance in terms of the F1 score. 
Notice that for WISE, Gaia, LINEAR, and ASAS, the RF achieves a higher recall score than the neural-based models. 
However, the RNNs models obtain a higher precision than the RF (except for the PLSTM trained on CSS data). We analyze these results in detail in Appendix \ref{appendix:cm}.
\\\\
We compared the performance of our models against other state-of-the-art deep learning classifiers.  \citealt{becker2020scalable} applied a Gated Recurrent Unit (GRU \citealt{cho2014learning}) over the unfolded lightcurves of WISE, OGLE, and Gaia. As shown in Table \ref{tab:summary}, our combination of LSTM and PLSTM models obtain a higher F1-score, outperforming Becker's results. 
\\\\
For the LINEAR and ASAS datasets, we compare our model against the recurrent autoencoder trained over folded lightcurves by \citealt{naul2018recurrent}. For the best configuration according to Table \ref{tab:summary}, we obtained an accuracy of 0.903 $\pm$ 0.007 for LINEAR and 0.971 $\pm$ 0.002 for ASAS. \citealt{naul2018recurrent} obtained an accuracy of 0.988 $\pm$ 0.003 for LINEAR, and 0.971 $\pm$ 0.006 for ASAS using folded lightcurves. However, our approach does not depend on folding the lightcurves, saving the computation time of calculating the periods. \citealt{naul2018recurrent} also report their accuracy over unfolded lightcurves, decreasing to 0.781 for LINEAR, significantly lower than our results. They did not explicitly report their accuracy for ASAS. 
\\\\
For the MACHO dataset, \citealt{zorich2020streaming} reached an F1 score of 0.86 by using a RF with incremental features over two bands. They also obtain an F1 score of 0.91 using an RF trained on FATS features \citep{nun2015fats}. Our model outperforms both results by obtaining an F1 score of 0.921 $\pm$ 0.017 even though we use only the \textit{B} band.
\begin{table*}
    \centering
    \input{Tables/summary_table}
    \caption{Offline table corresponding to the model evaluation using the entire lightcurve. The first and second columns are associated with the dataset and the classification model, respectively. The third column denotes the normalization technique associated with the best cross-validated metrics on columns 4, 5, and 6. We also include the available scores from \protect\citealt{becker2020scalable}, which is the current state of the art for WISE, Gaia, and OGLE.}
    \label{tab:summary}
\end{table*}
\\\\
Taking advantage of the PLSTM properties, we have shown that the L+P model improves the offline classification of lightcurves even when the single PLSTM performs worse than the other models (i.e., LSTM and RF). Our model is able to classify using unfolded lightcurves, which avoids the computational cost of calculating the period to fold the lightcurves, and the number of points needed to perform classification on a given lightcurve.
\subsection{Streaming Evaluation}\label{streaming}
\begin{figure}
    \centering
    \includegraphics[scale=0.6]{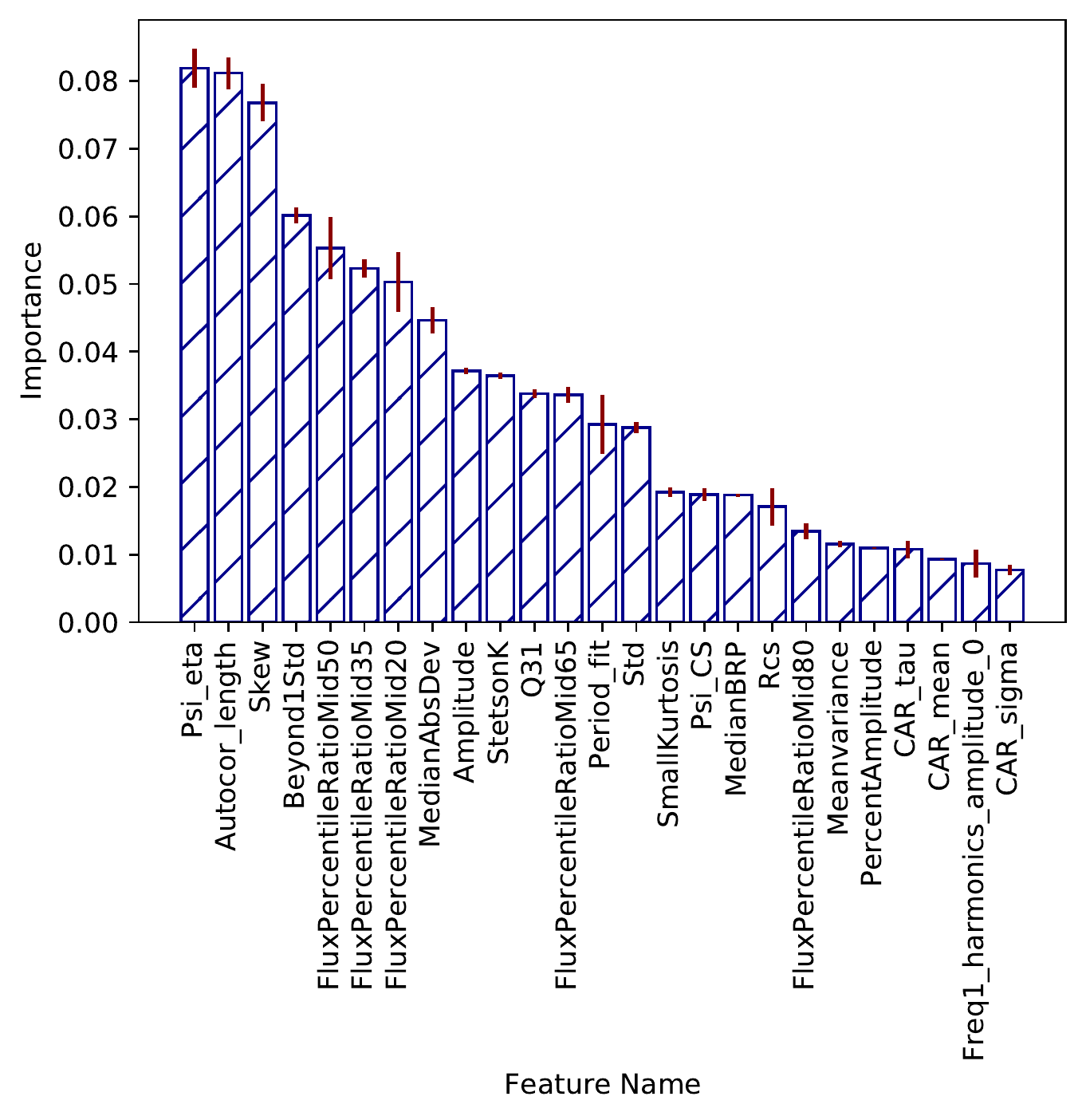}
    \caption{LINEAR Feature importance according to the Gini-score of the Random Forest classifier.}
    \label{fig:linear_importance}
\end{figure}

\begin{figure}
    \centering
    \includegraphics[scale=0.6]{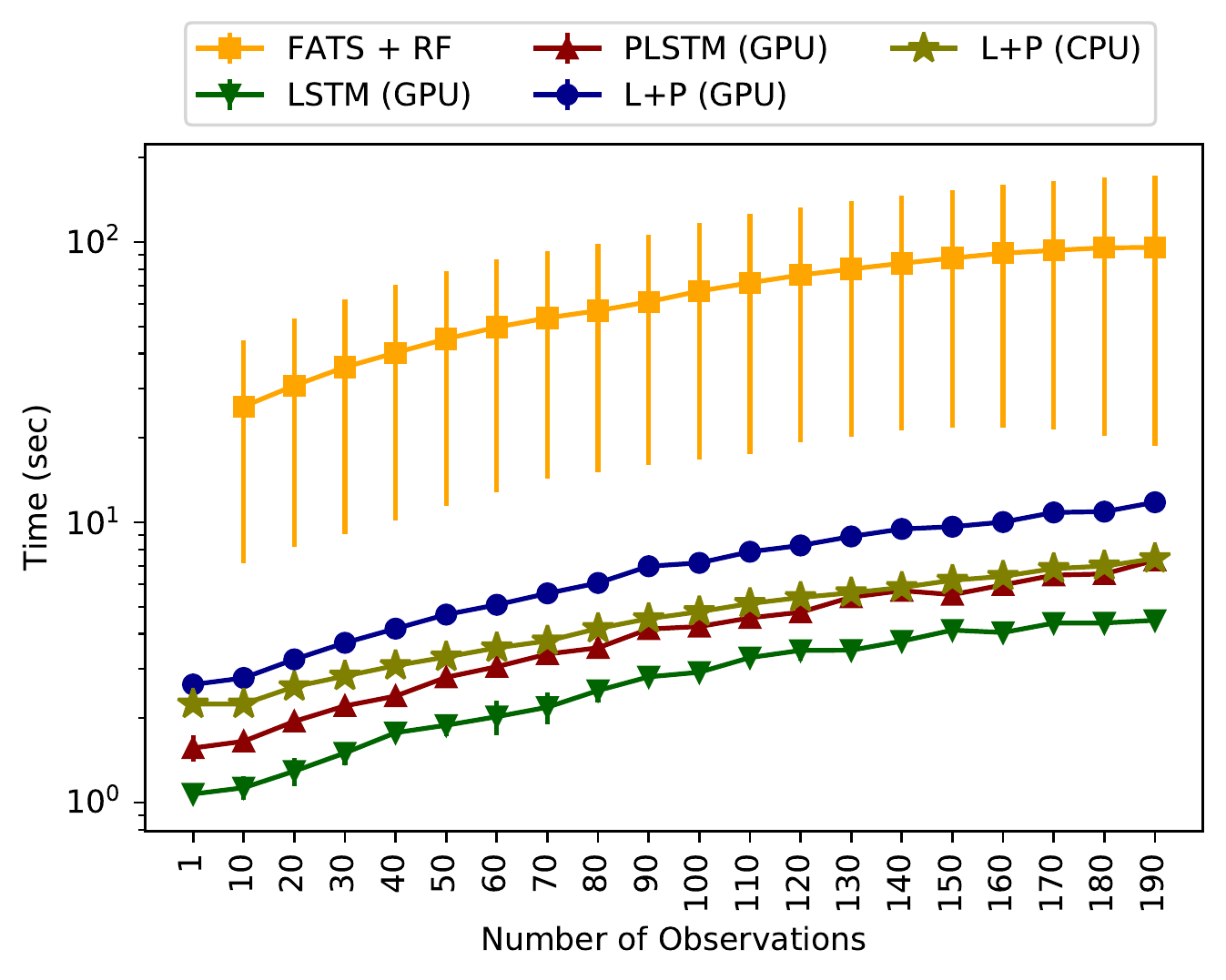}
    \caption{Testing run-times in function of the number of observations among different surveys.}
    \label{fig:pred_times}
\end{figure}

\begin{figure*}
    \centering
    \includegraphics[scale=0.55]{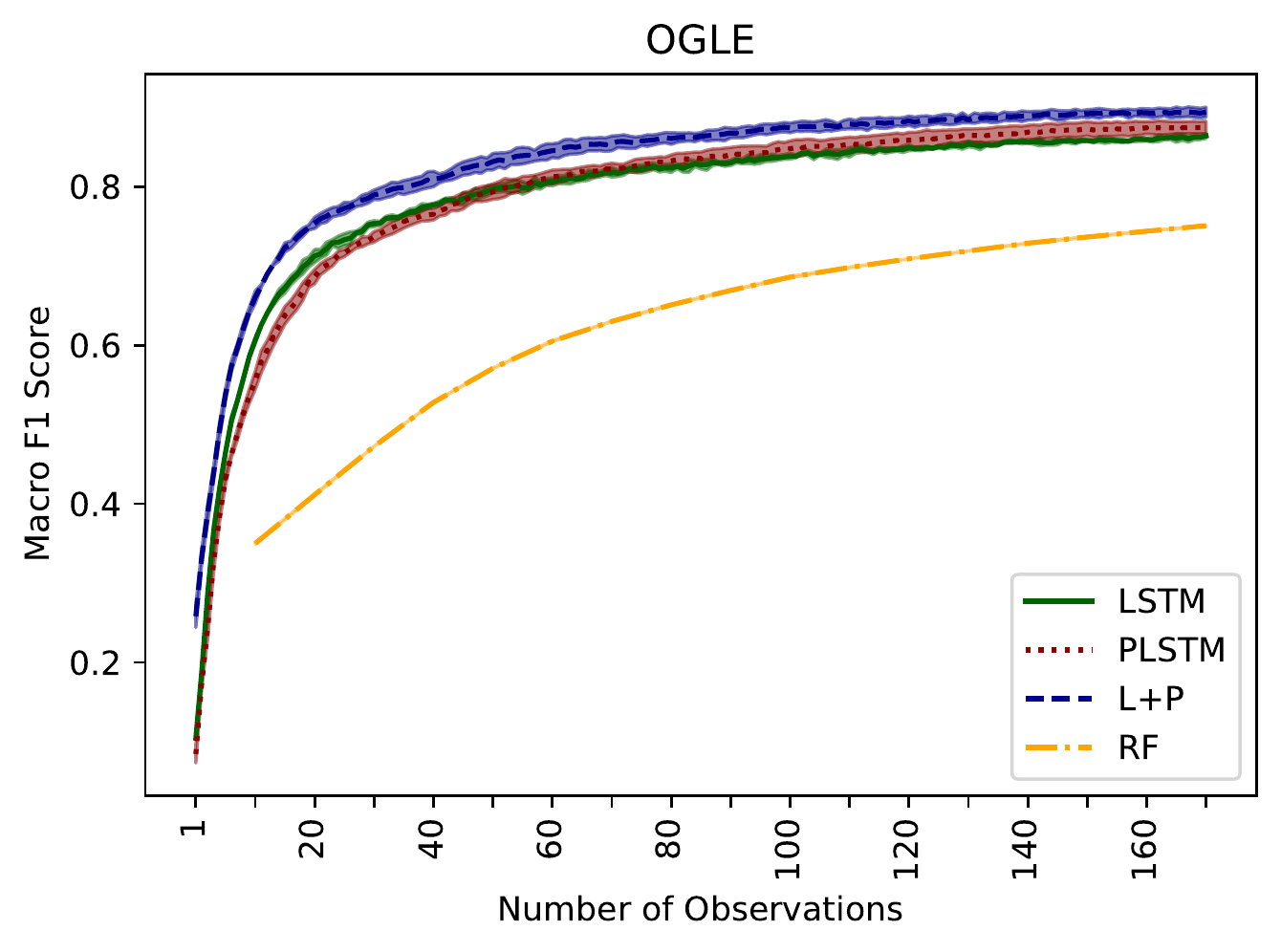}\hspace{1mm}
    \includegraphics[scale=0.55]{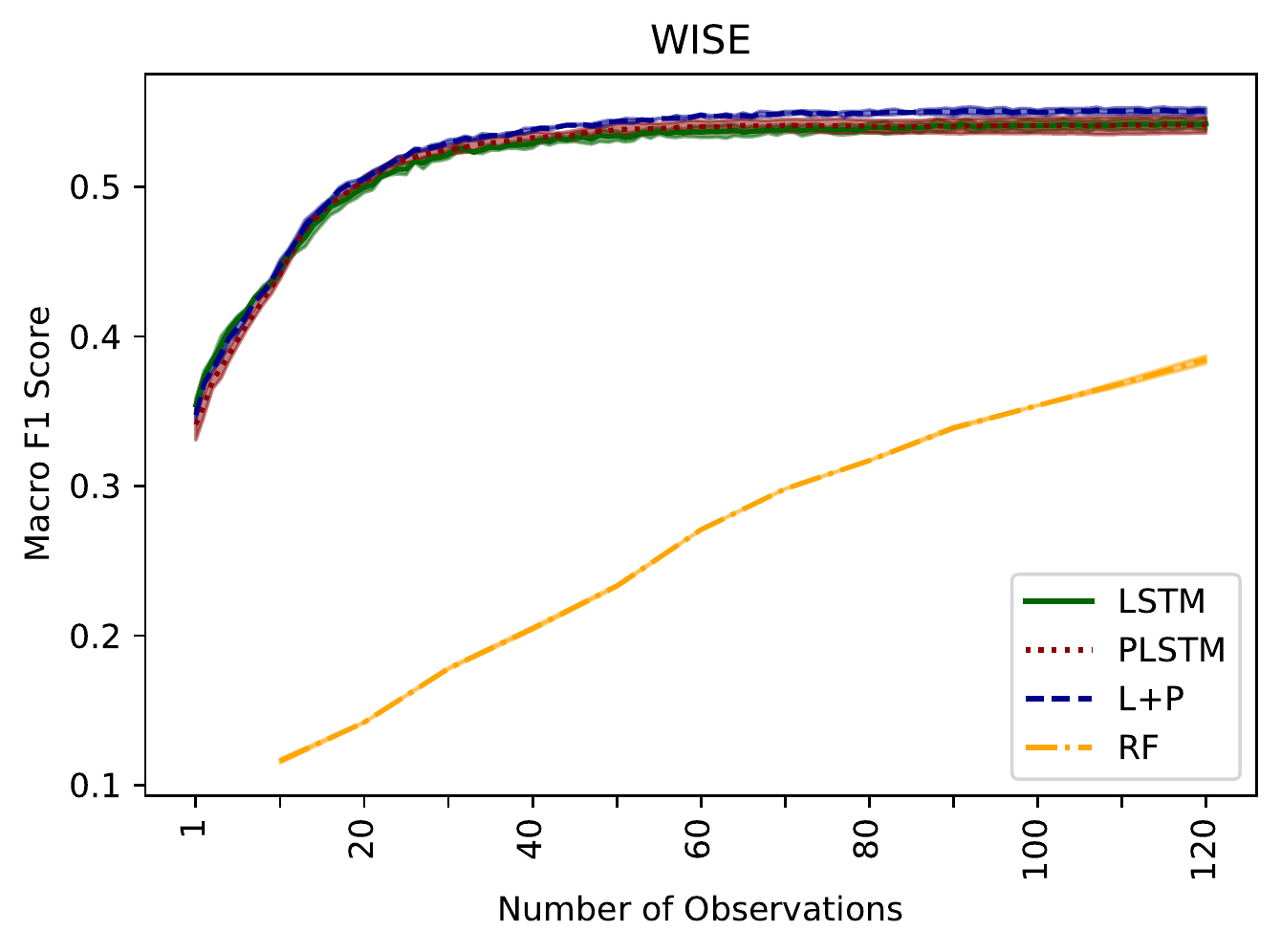}\\
    \includegraphics[scale=0.55]{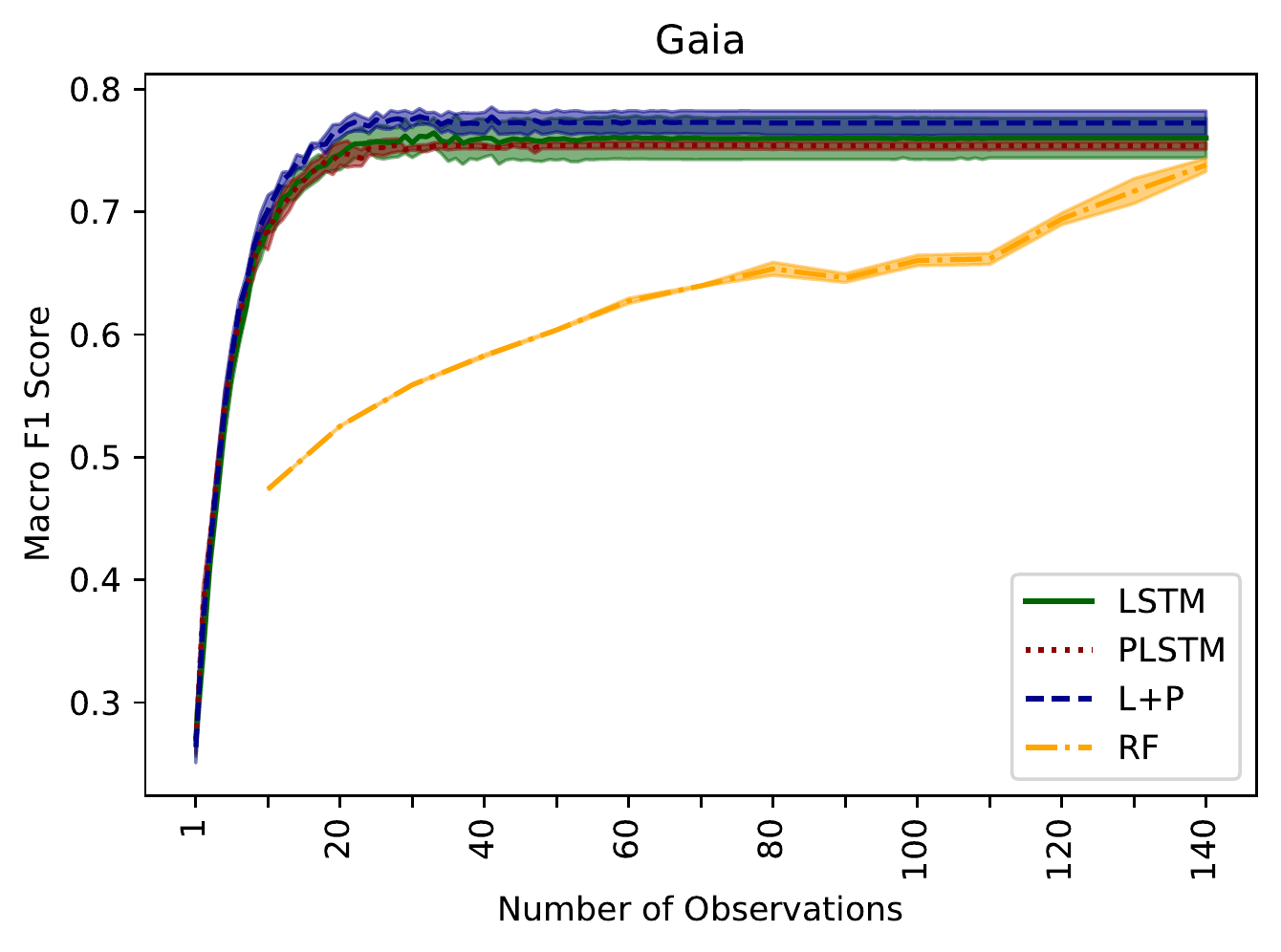}\hspace{1mm}
    \includegraphics[scale=0.55]{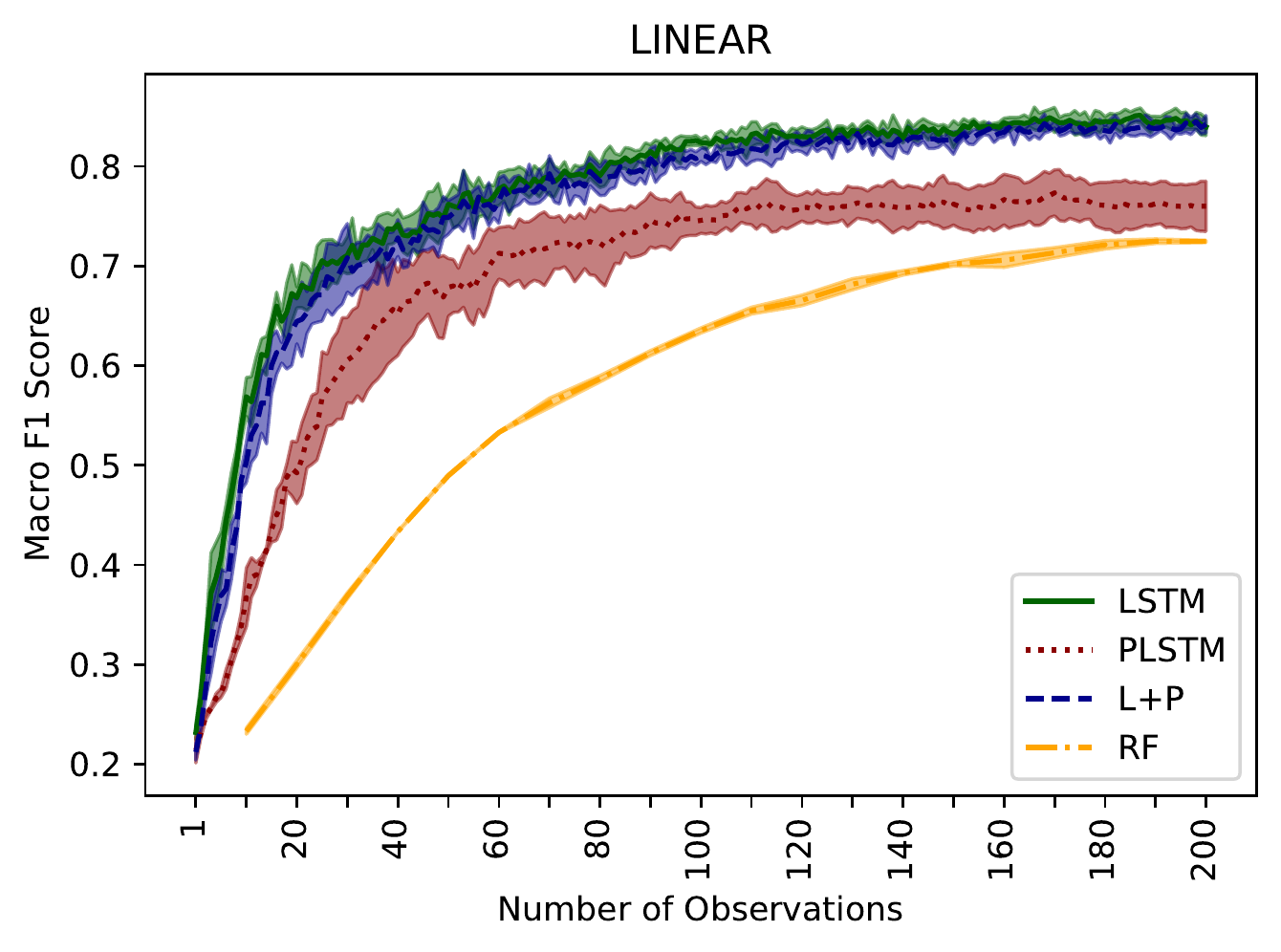}\\
    \includegraphics[scale=0.55]{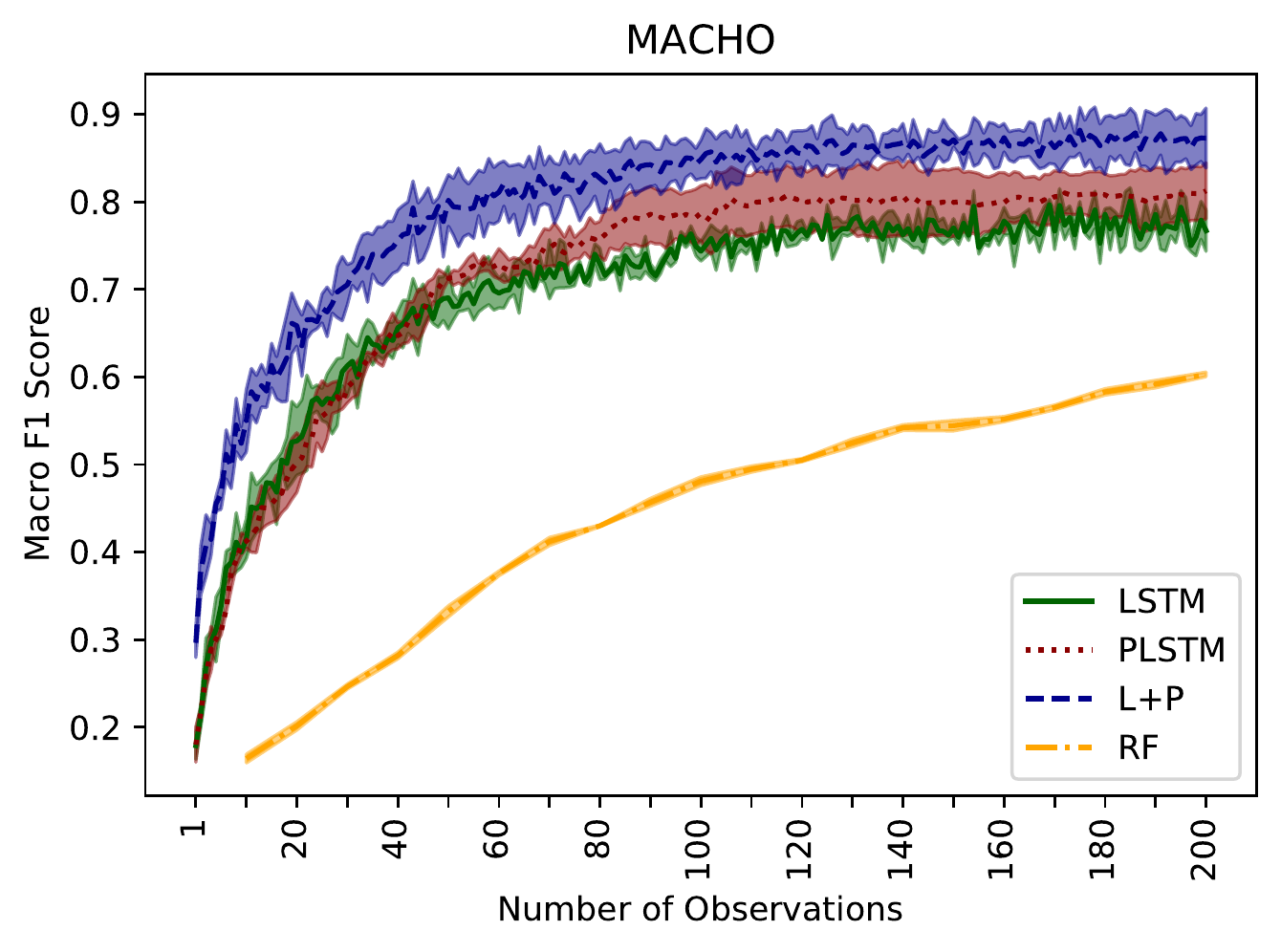}\hspace{1mm}
    \includegraphics[scale=0.55]{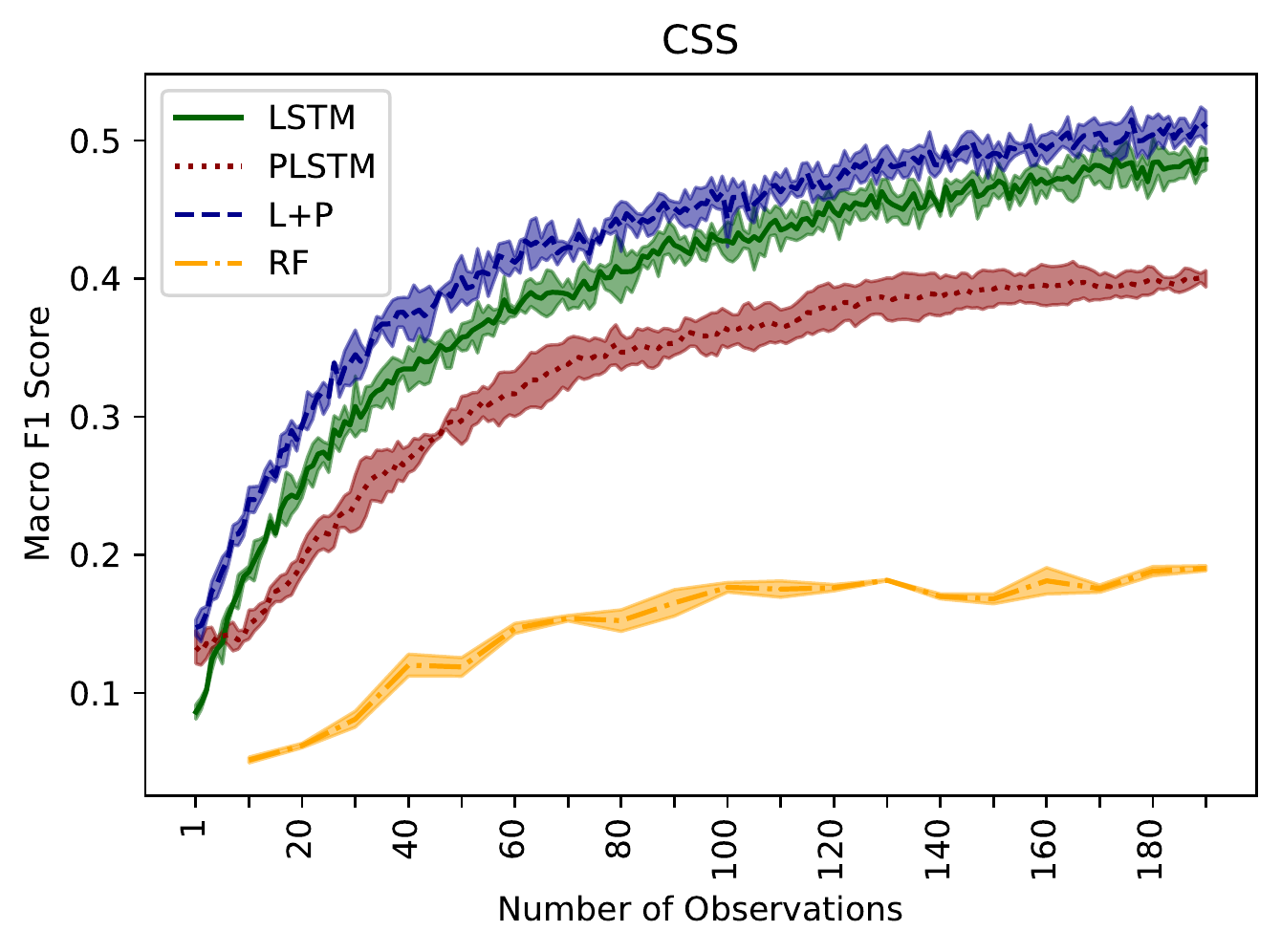}\\
    \includegraphics[scale=0.55]{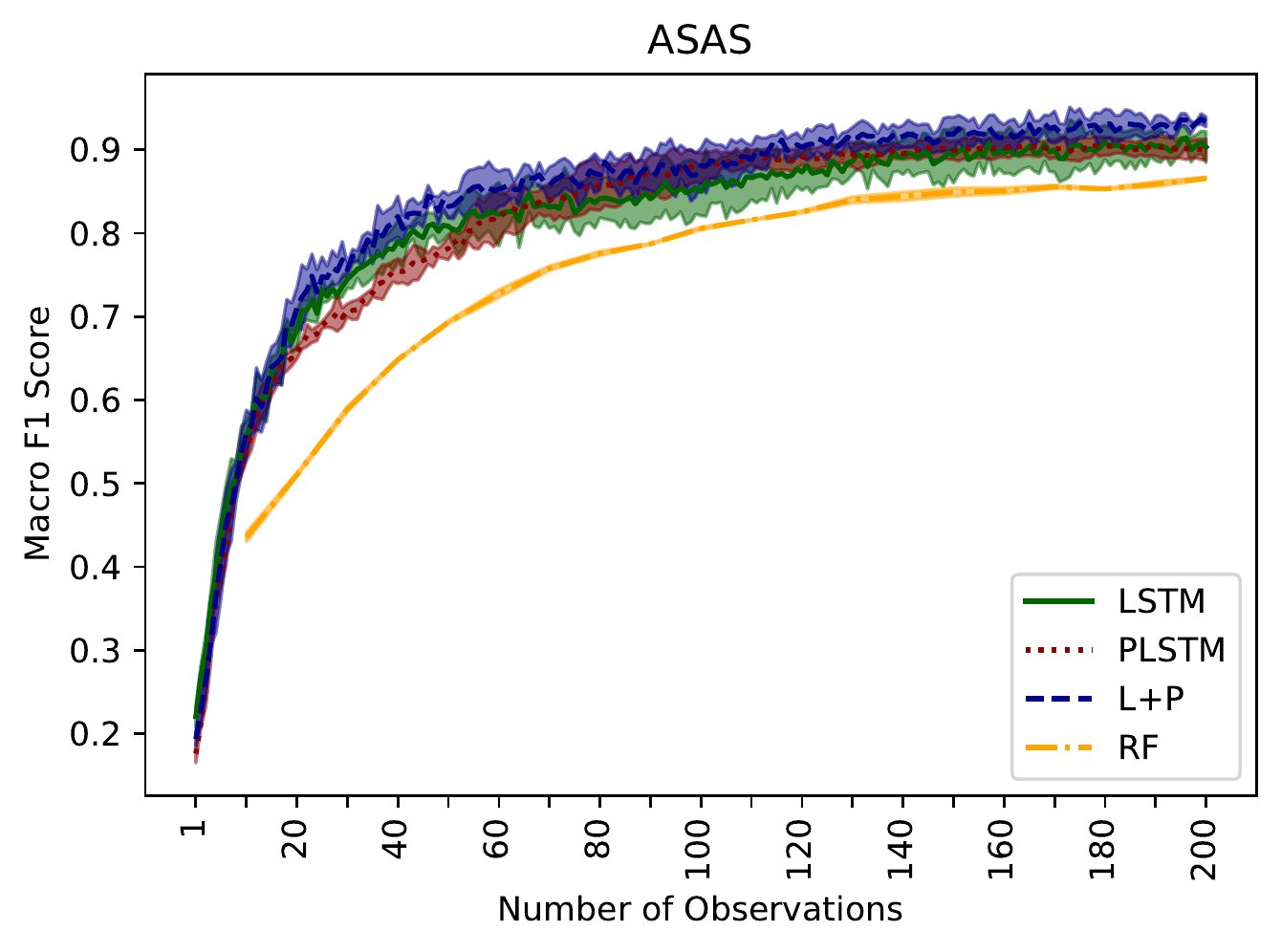}
    \caption{Streaming evaluation of the Random Forest (RF), Long Short Term Memory (LSTM), Phased LSTM (PLSTM), and the combination of both recurrent units (L+P) for each dataset. All curves show the mean and standard deviation of the macro F1 score as a function of the number of observations.}
    \label{fig:online_evaluation}
\end{figure*}

In Section \ref{offline}, we evaluate our models using all lightcurve observations. However, when analyzing real-world data, we require classifying sources online (\citealt{borne2007machine};
\citealt{saha2014antares};
\citealt{borne2009scientific}; \citealt{forster2020automatic}). To assess our models' quality in this scenario, we evaluate them in terms of the number of observations presented in the lightcurve.
\\\\
Figure \ref{fig:online_evaluation} shows the F1-score of the RF, LSTM, PLSTM, and L+P models for each dataset in terms of the number of points used for the classification. As we mention on Section \ref{sec:methodology}, the classifiers were trained using all observations in the lightcurves subsamples, while this evaluation is performed in terms of the number of points of the lightcurves. As the number of observations increases, the F1 scores of neural-based models increase faster than those for the RF. We hypothesize that some pre-calculated features could be more affected by short-length lightcurves than the representations made by RNNs. We study this premise in Fig. \ref{fig:linear_importance}, where we show the features importance according to the RF model for the LINEAR dataset. Attributes related to position, dispersion, and central tendency are important to separate classes in this case. These features  depend on the number of observations. Conversely, RNNs propagate the classification error at each time step during training, forcing the network to discover better separability of the classes in early epochs.
\\\\
Our L+P model takes advantage of the LSTM and PLSTM units, improving the scores at different instants of the lightcurve. In most cases, the LSTM and PLSTM models achieve similar results (with the exception of LINEAR and CSS, where the LSTM model significantly outperforms the PLSTM model). The L+P improves or equals the F1-score of the best models by averaging the probabilities of both recurrent architectures.
\\\\
For the RF, most of the features need to be re-calculated when a new sample arrives, which is time-consuming compared to the RNNs. Figure \ref{fig:pred_times} shows the mean and standard deviation of the prediction times for the feature-based classifier and RNNs tested on the entire sequence across all datasets. The execution time of the RFs for classifying lightcurves grows faster than RNNs because it depends on the feature extraction process. For example, the period estimation is $O(L\log L)$ and the autocorrelation function is $\mathcal{O}(L^2)$, where $L$ is the number of observations. In contrast, RNNs only need to update the last hidden state, $\mathcal{O}(1)$ for a single observation and $\mathcal{O}(L)$ for the entire sequence. 

\section{Conclusions}\label{sec:conclusions}
In this work, we addressed the problem of the automatic classification of unfolded real lightcurves. We introduced a deep learning model (L+P) that combines the output probabilities from an LSTM and PLSTM classifiers. Empirical evaluations on seven single-band catalogs, including periodic and nonperiodic astronomical objects,  demonstrated the L+P model outperforms models trained on features and state-of-the-art classifiers. 
\\\\
We tested our proposed model both using the whole lightcurve and incrementally in time. The L+P architecture obtained the best F1 score among all datasets in both scenarios. Moreover, the model turned out to be an excellent alternative to classify short-length lightcurves where the number of observations is less than 20. The cumulative loss used to update weights forces the network to accurately predict all time steps, not just in the last observation.
\\\\
Recurrent neural networks need only the last hidden state to predict new observations. The hidden states save features of the lightcurve up to the last observed point. The update process is O(1), which is ideal for streaming scenarios, unlike the RF that needs to recalculate features using all time steps. 
\\\\
Since the neural-based models learn representations of the unprocessed raw data, we can capture features without restricting the domain using prior knowledge, such as the object's periodicity. However, the optimization process may be suboptimal when we have a low number of samples to fit the model parameters. Future works will focus on providing more information to the network by adding, for example, stellar coordinates or multiple bands during training.
\\\\
We have proved the robustness of our architecture on different kinds of stars and astronomical catalogs. Our classifier works on unfolded periodic and nonperiodic lightcurves, achieving comparable or better-than results than other state-of-the-art approaches. We will continue to delve into these neural architectures facing the new generation of telescopes and their new real-world applications.

\section*{Acknowledgements}
The authors acknowledge support from ANID Millennium Science Initiative ICN12 009, awarded to the Millennium Institute of Astrophysics; FONDECYT Regular 1171678 (P.E.); FONDECYT Initiation Nº 11191130 (G.C.V.); This work has been possible thanks to the use of AWS-NLHPC credits. Powered@NLHPC: This research was partially supported by the supercomputing infrastructure of the NLHPC (ECM-02).

\section*{Data Availability}
All data can be found at \url{https://drive.google.com/drive/folders/1m2fXqn25LYSyG5jEbpM3Yfdbpx9EQCDo?usp=sharing} or downloading directly as indicated in \url{https://github.com/cridonoso/plstm_tf2}.


\clearpage
\bibliographystyle{mnras}
\bibliography{references} 




\appendix
\section{Confusion Matrices}\label{appendix:cm}
\begin{figure}
    \centering
    \includegraphics[scale = 0.2]{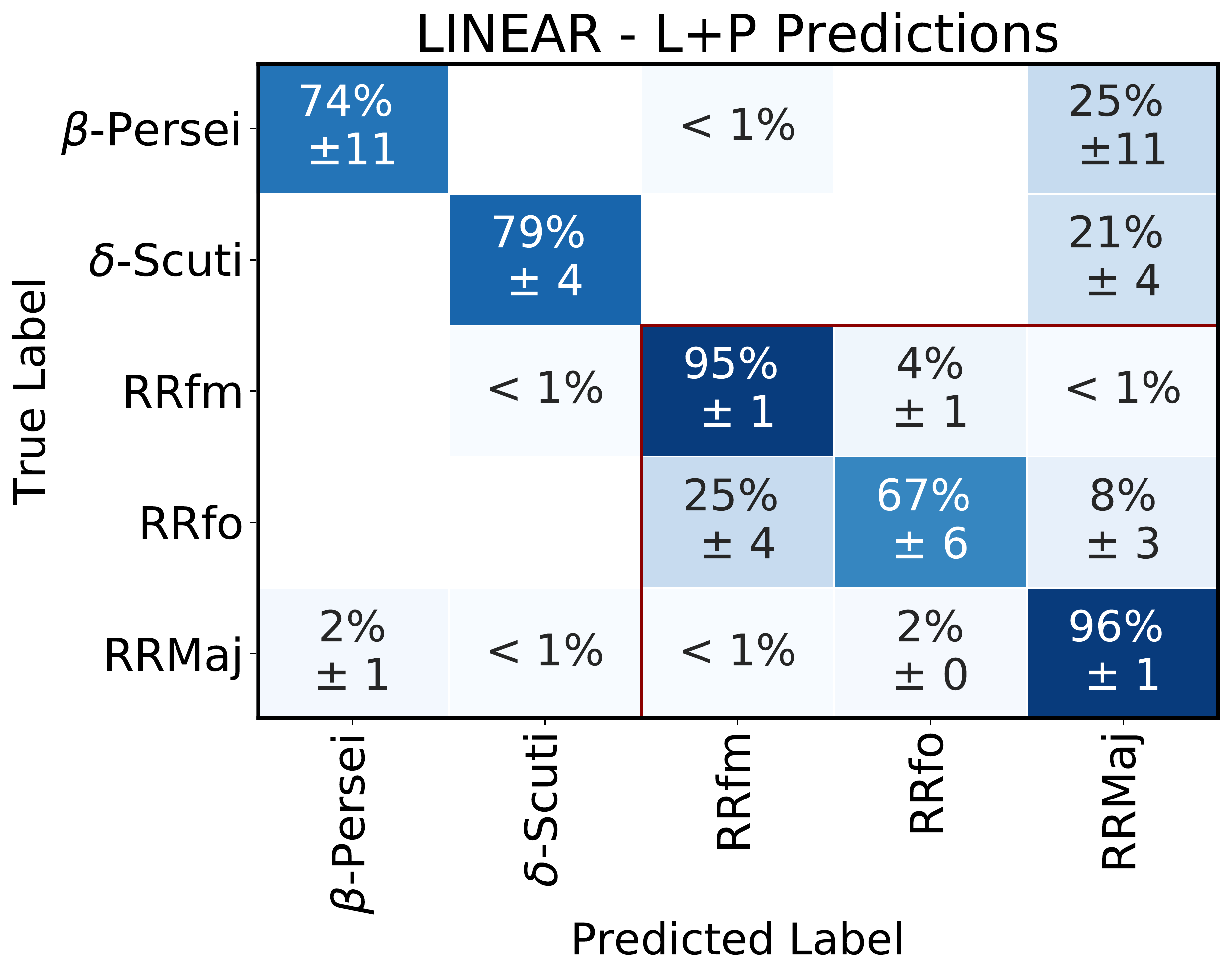}
    \includegraphics[scale = 0.2]{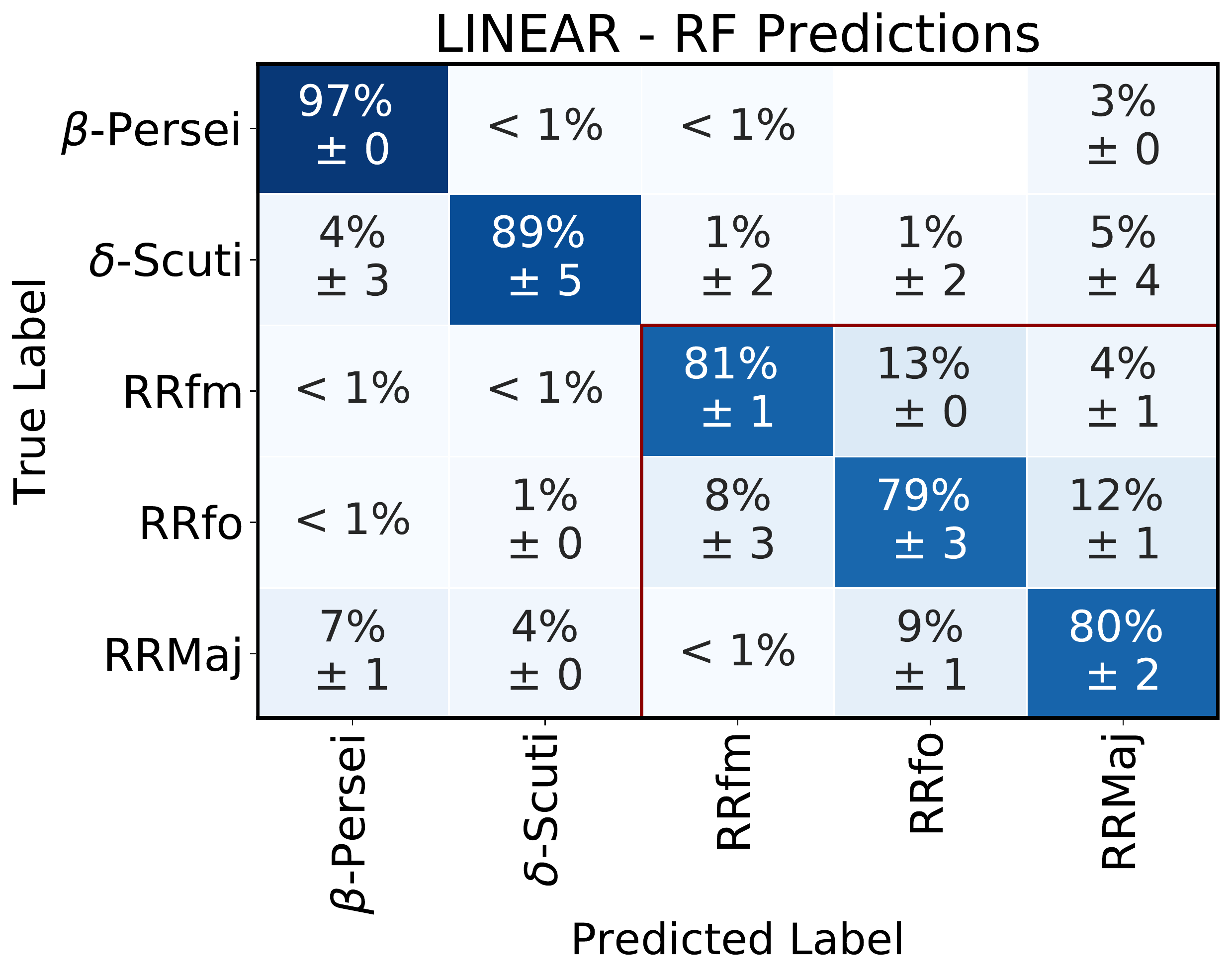}
    \caption{LINEAR confusion matrices for the predictions of the LSTM+PLSTM (L+P) and the Random Forest (RF) classifiers.}
    \label{cm:linear}
\end{figure}

\begin{figure}
    \centering
    \includegraphics[scale = 0.2]{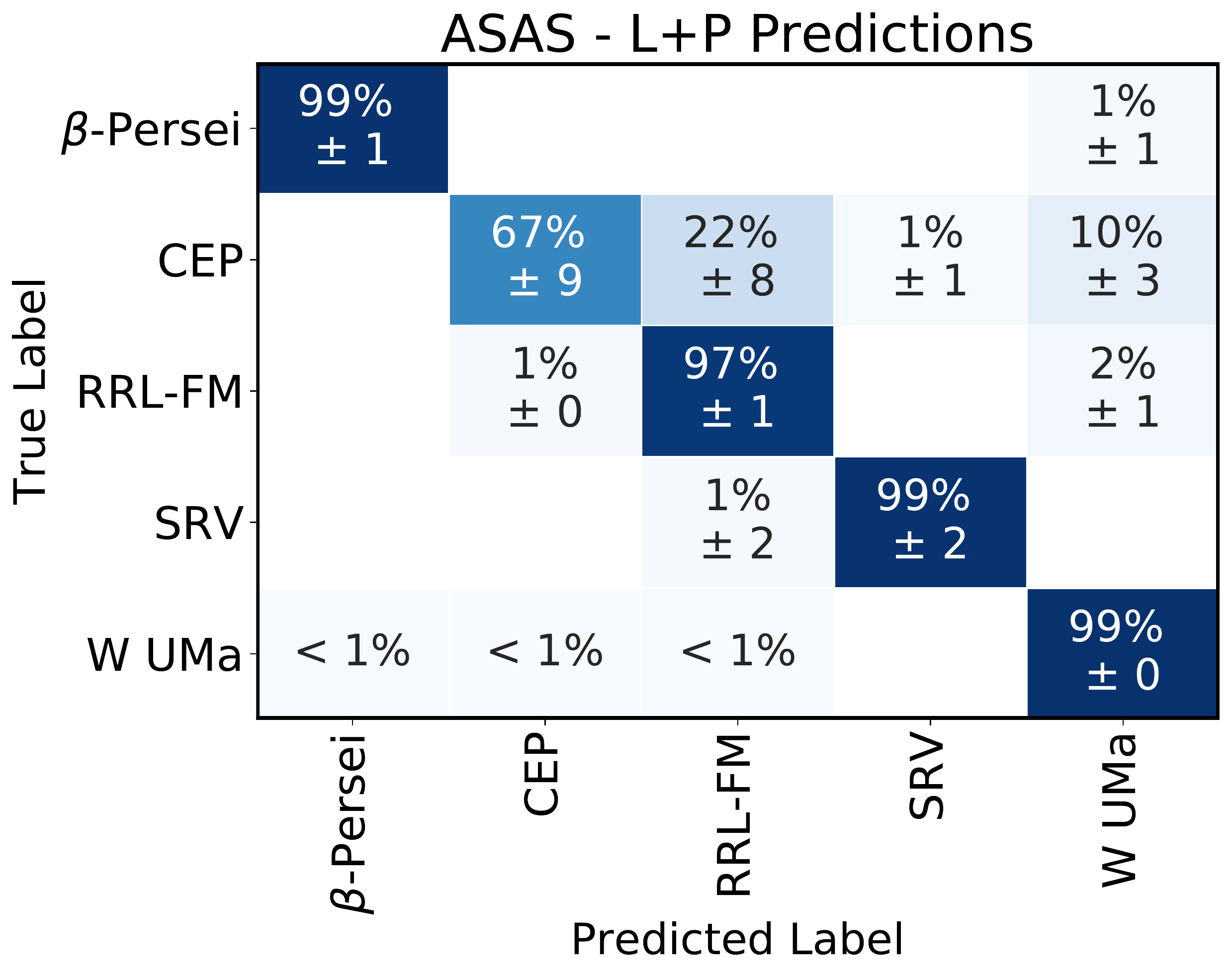}
    \includegraphics[scale = 0.2]{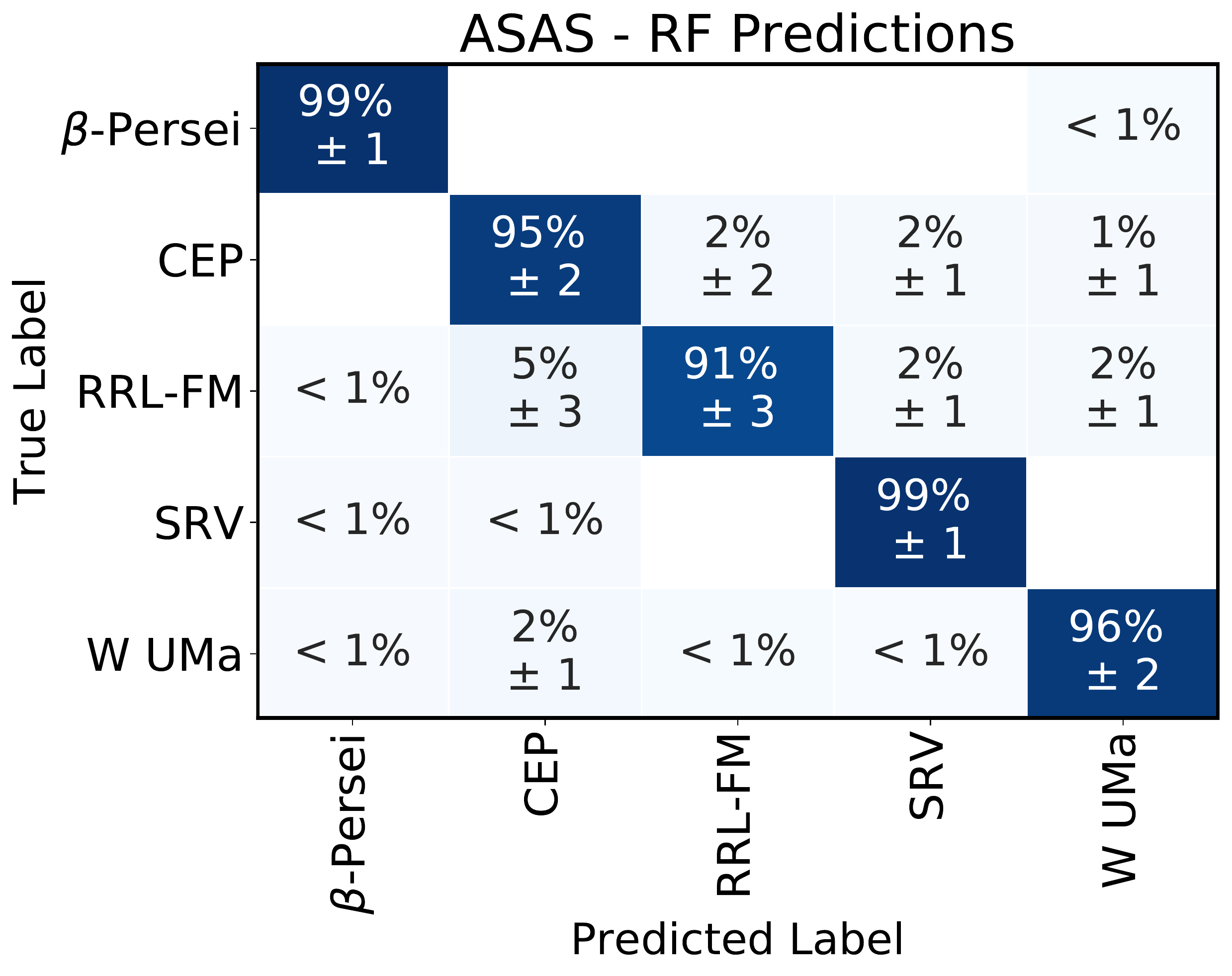}
    \caption{ASAS confusion matrices for the predictions of the LSTM+PLSTM (L+P) and the Random Forest (RF) classifiers.}
    \label{cm:asas}
\end{figure}

\begin{figure}
    \centering
    \includegraphics[scale = 0.35]{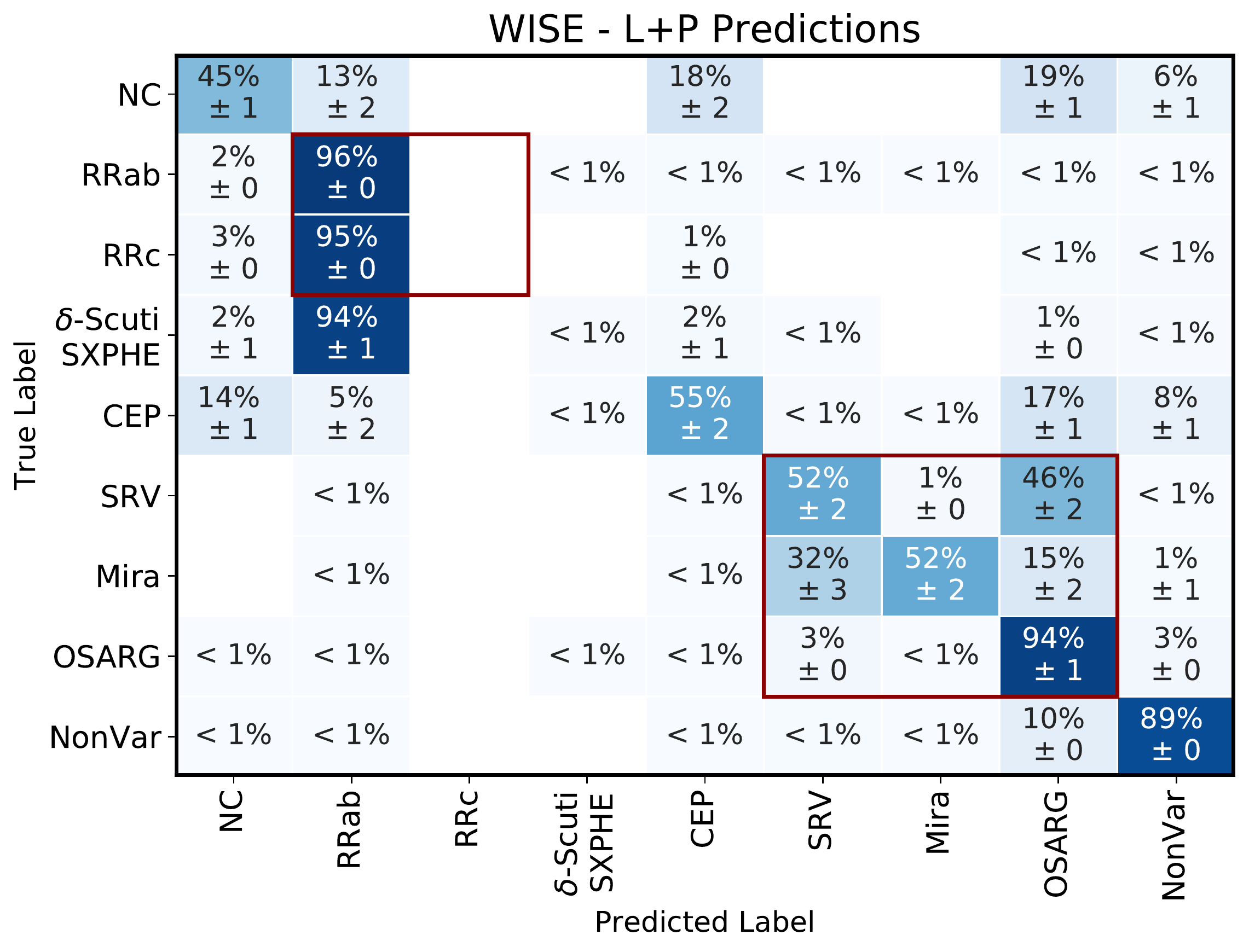}
    \includegraphics[scale = 0.35]{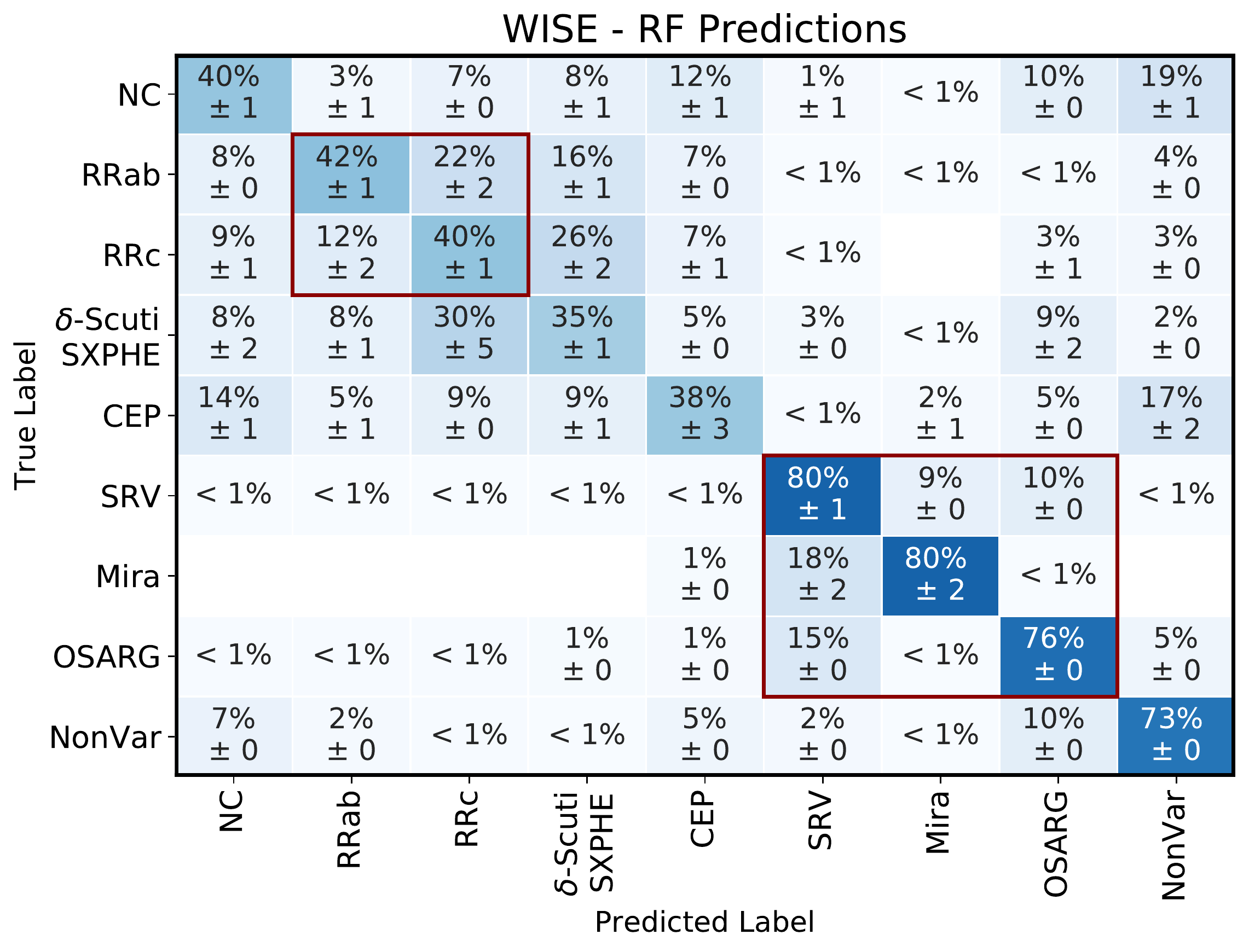}
    \caption{WISE confusion matrices for the predictions of the LSTM+PLSTM (L+P) and the Random Forest (RF) classifiers.}
    \label{cm:wise}
\end{figure}

\begin{figure}
    \centering
    \includegraphics[scale = 0.25]{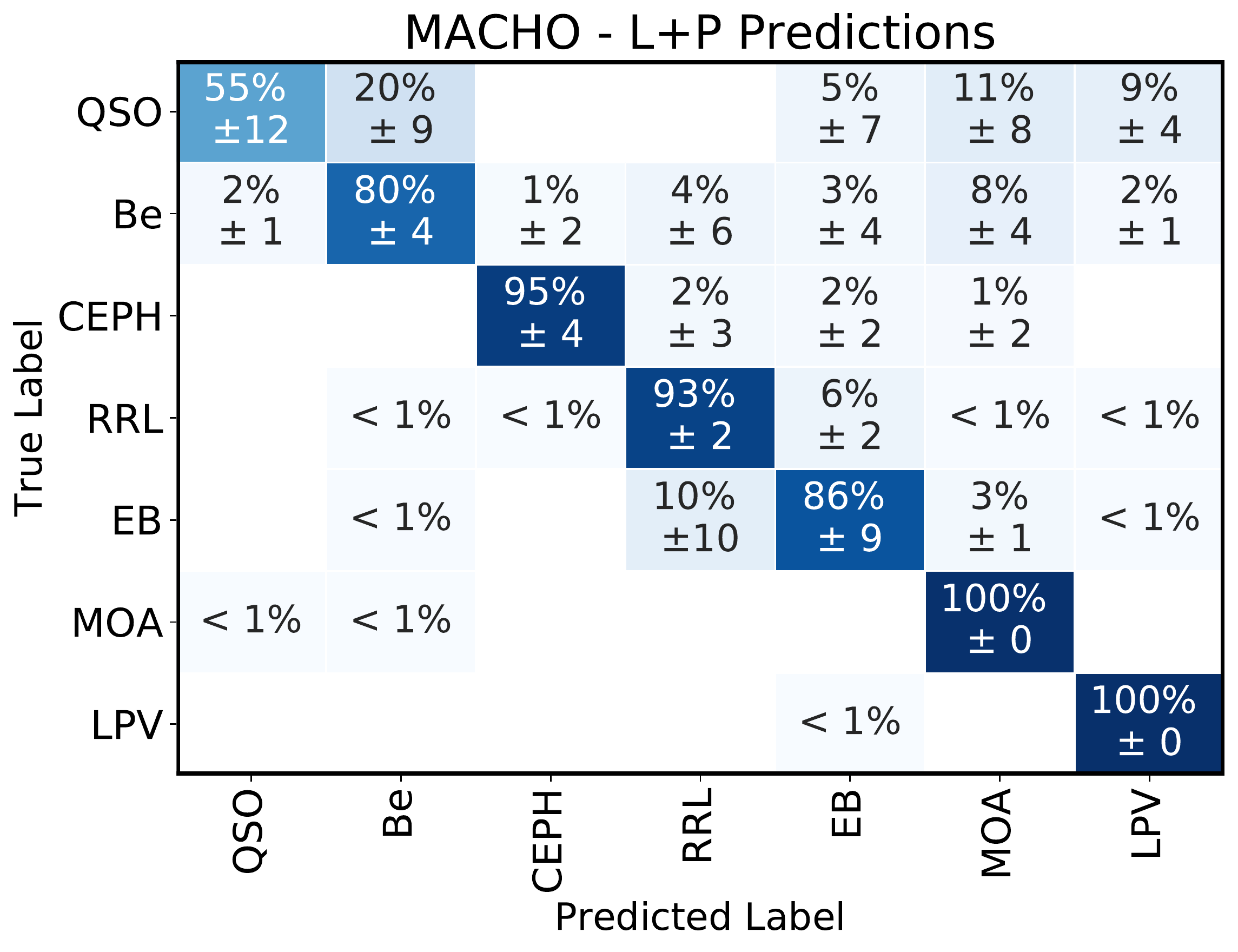}
    \includegraphics[scale = 0.25]{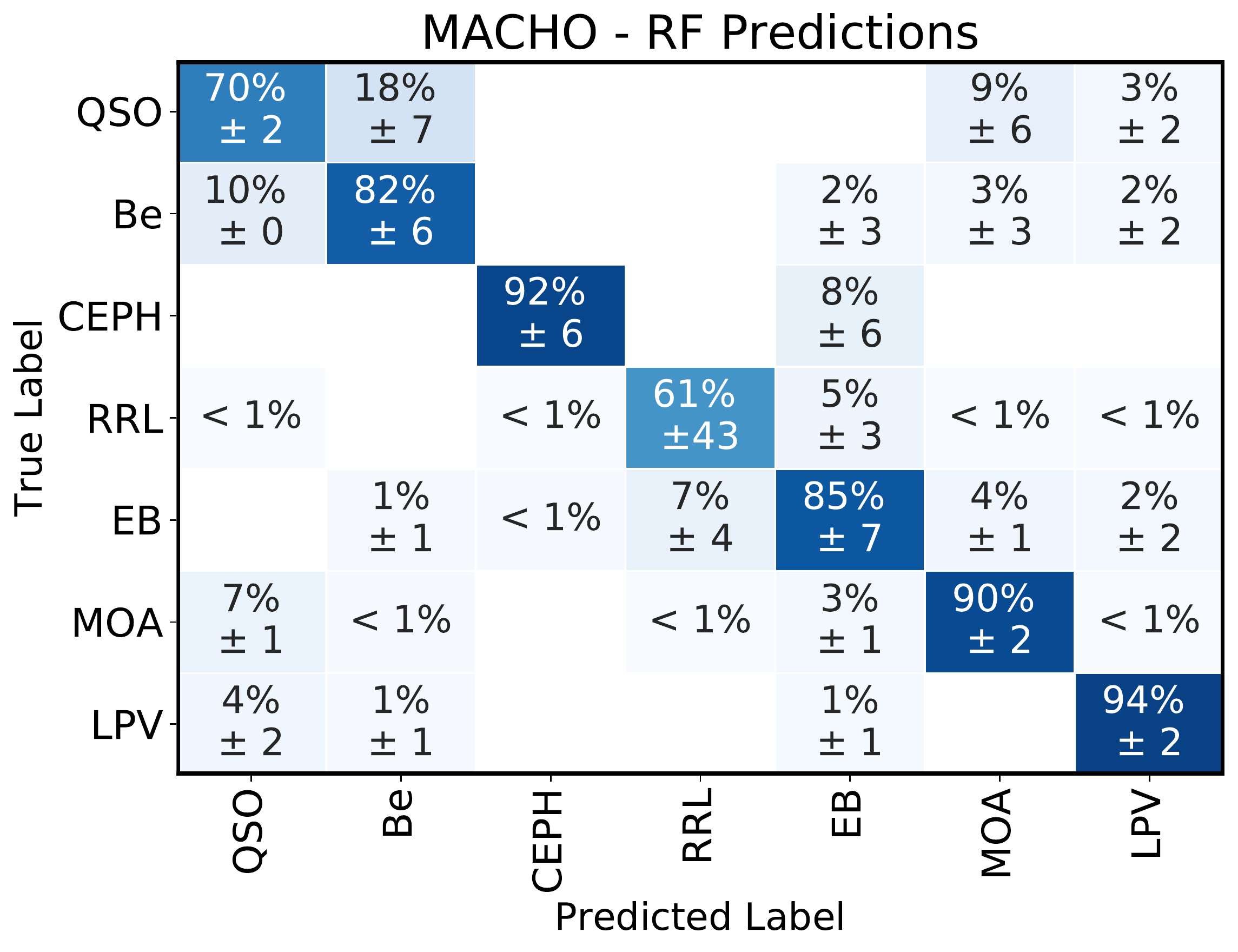}
    \caption{MACHO confusion matrices for the predictions of the LSTM+PLSTM (L+P) and the Random Forest (RF) classifiers.}
    \label{cm:macho}
\end{figure}

\begin{figure}
    \centering
    \includegraphics[scale = 0.3]{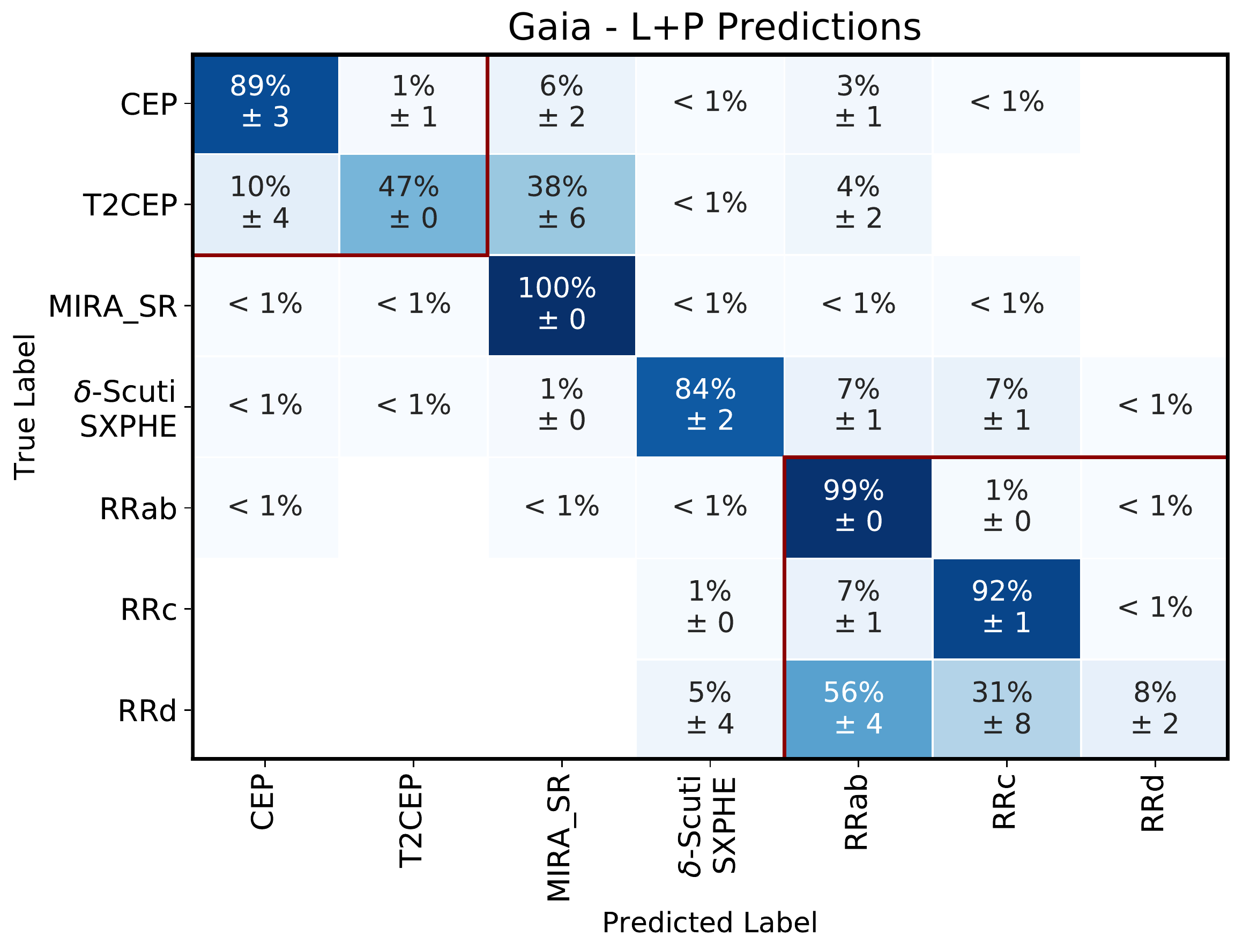}
    \includegraphics[scale = 0.3]{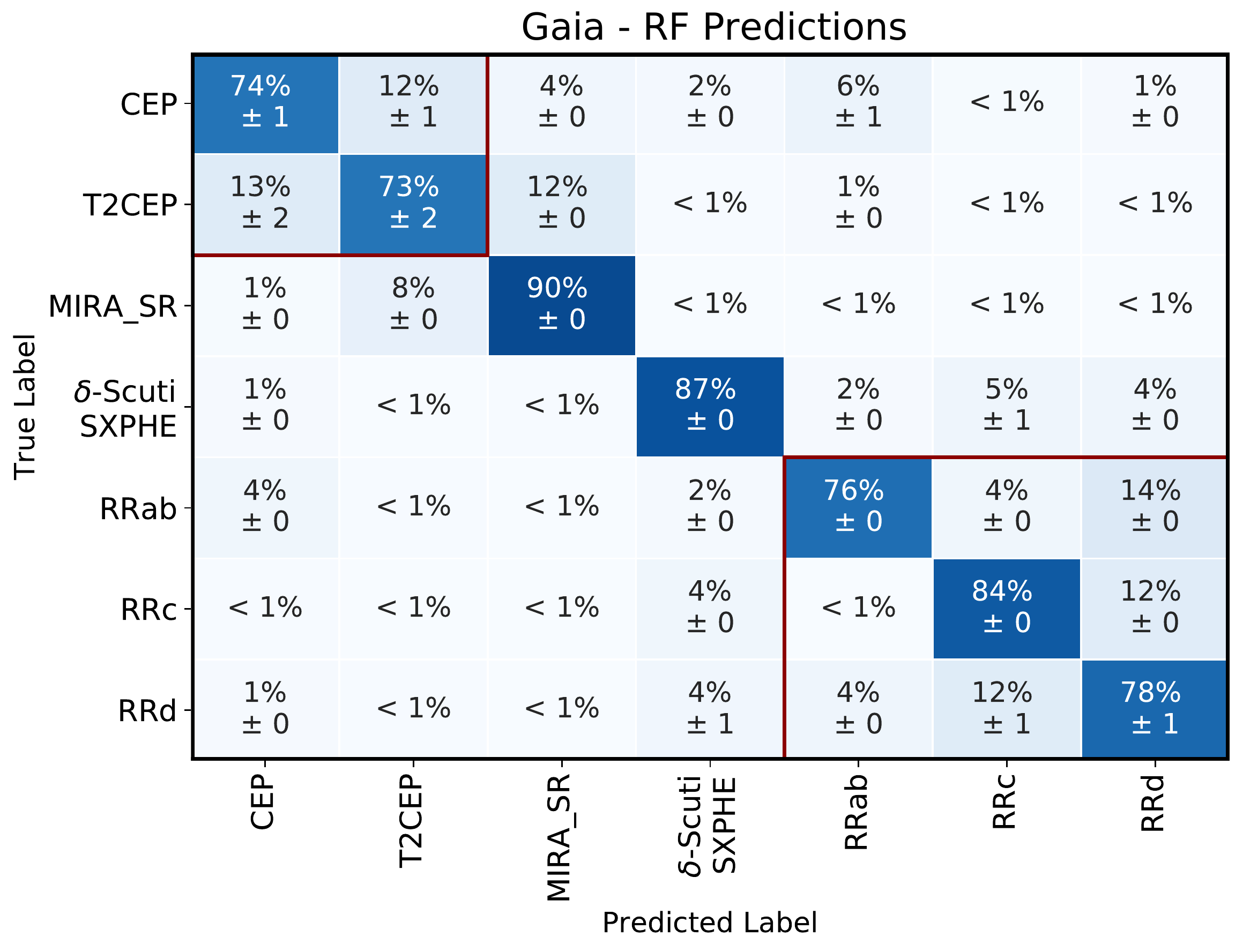}
    \caption{Gaia confusion matrices for the predictions of the LSTM+PLSTM (L+P) and the Random Forest (RF) classifiers.}
    \label{cm:gaia}
\end{figure}

\begin{figure}
    \centering
    \includegraphics[scale = 0.35]{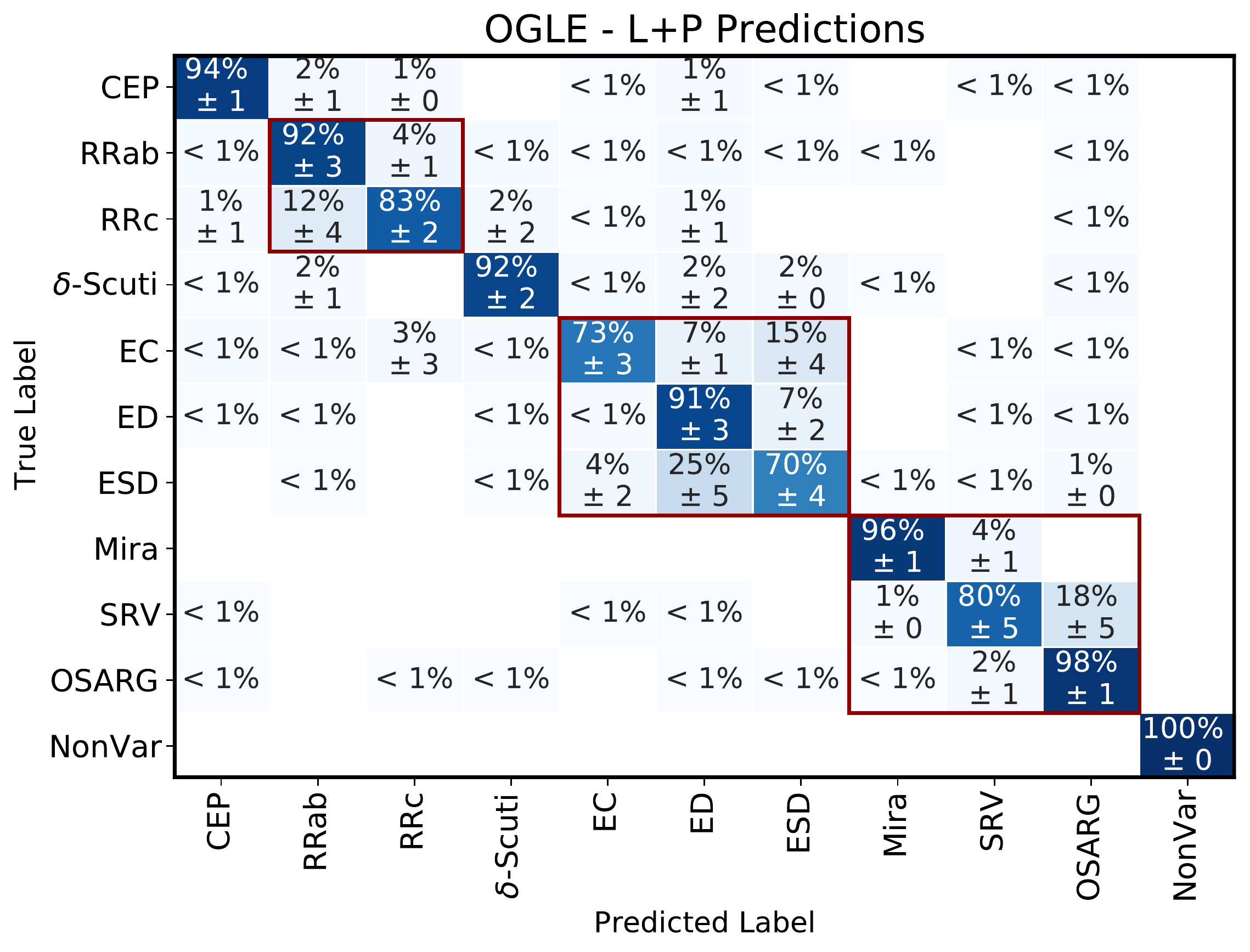}
    \includegraphics[scale = 0.35]{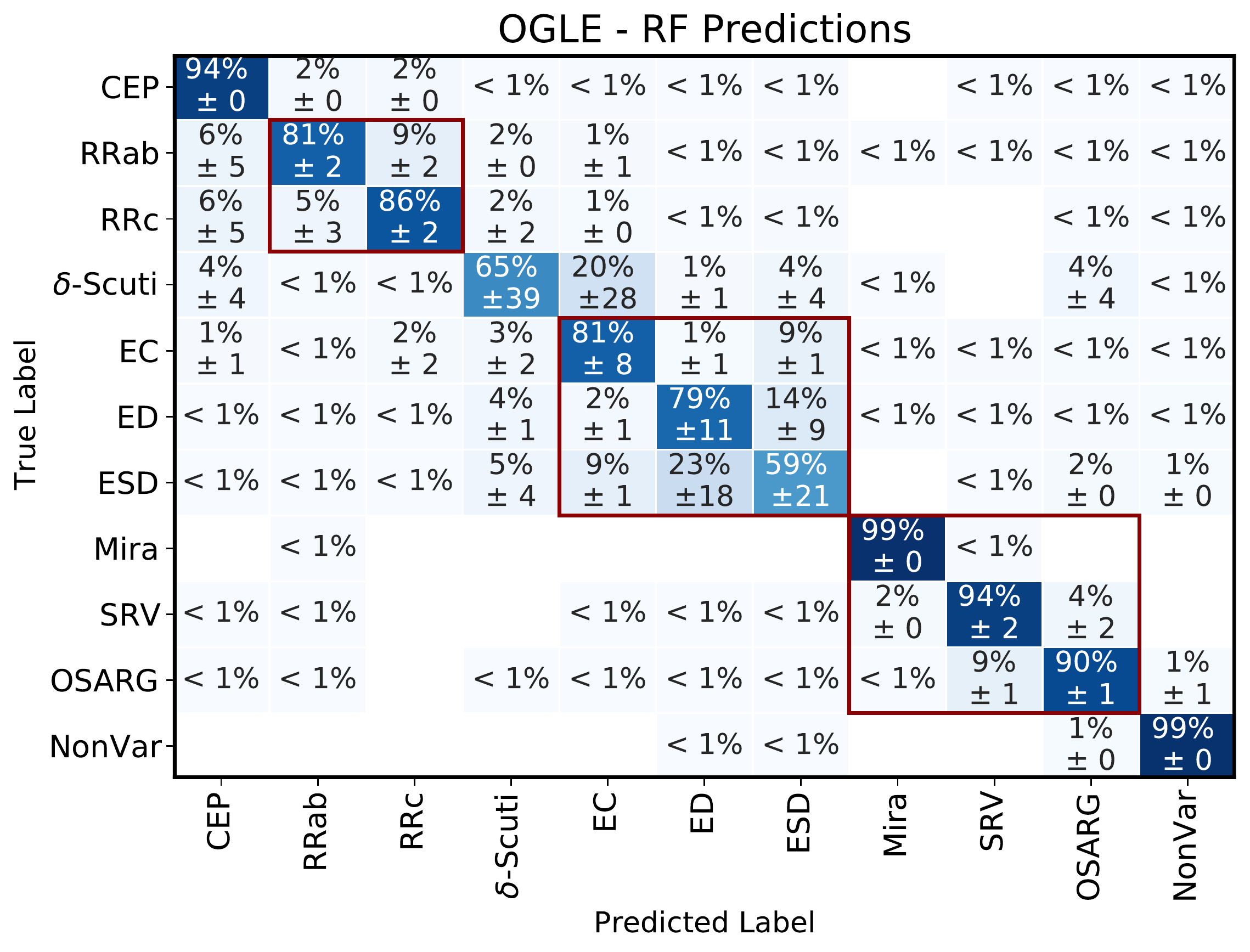}
    \caption{OGLE confusion matrices for the predictions of the LSTM+PLSTM (L+P) and the Random Forest (RF) classifiers.}
    \label{cm:ogle}
\end{figure}

\begin{figure*}
    \centering
    \includegraphics[scale = 0.69]{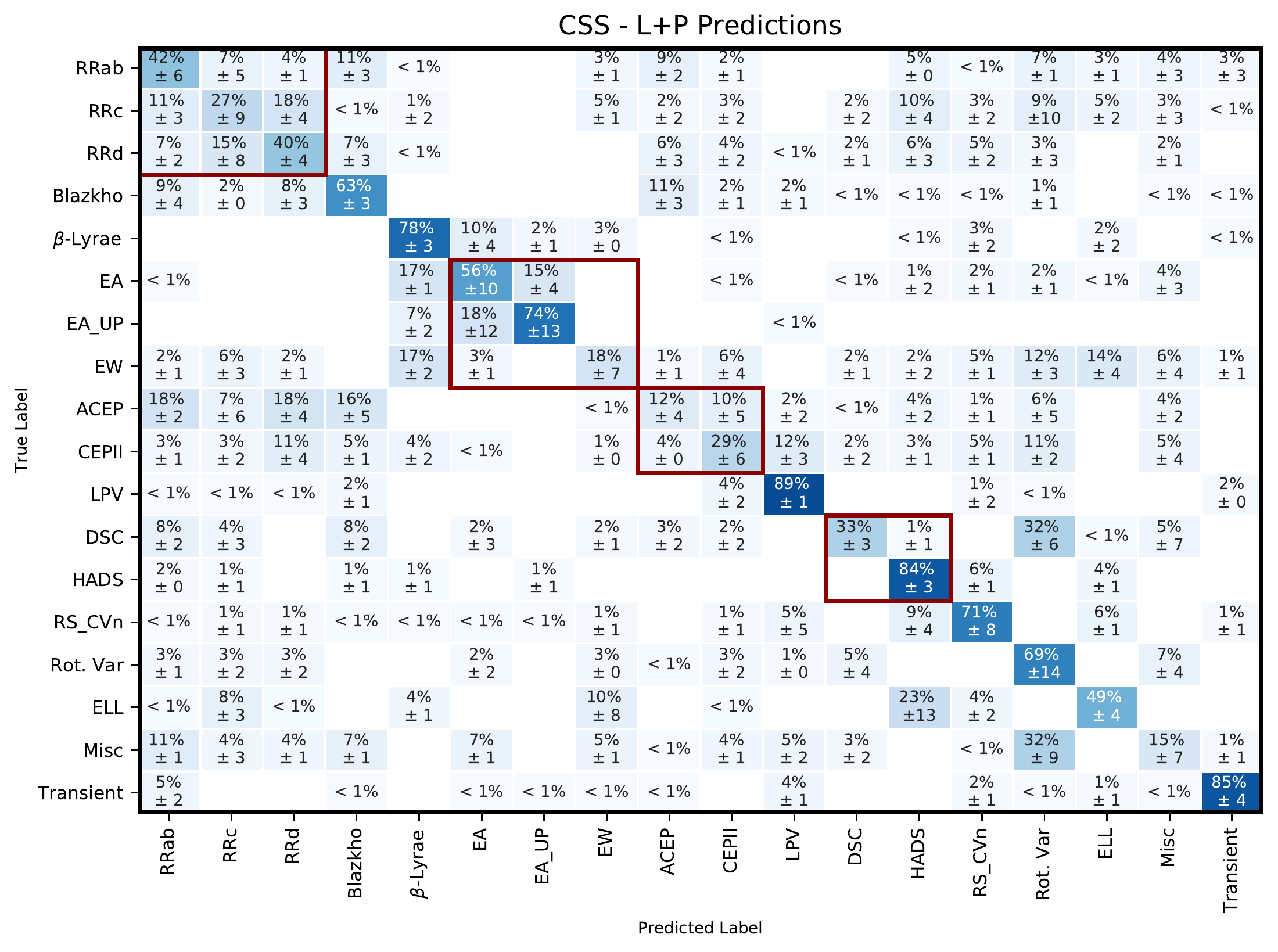}
    \includegraphics[scale = 0.69]{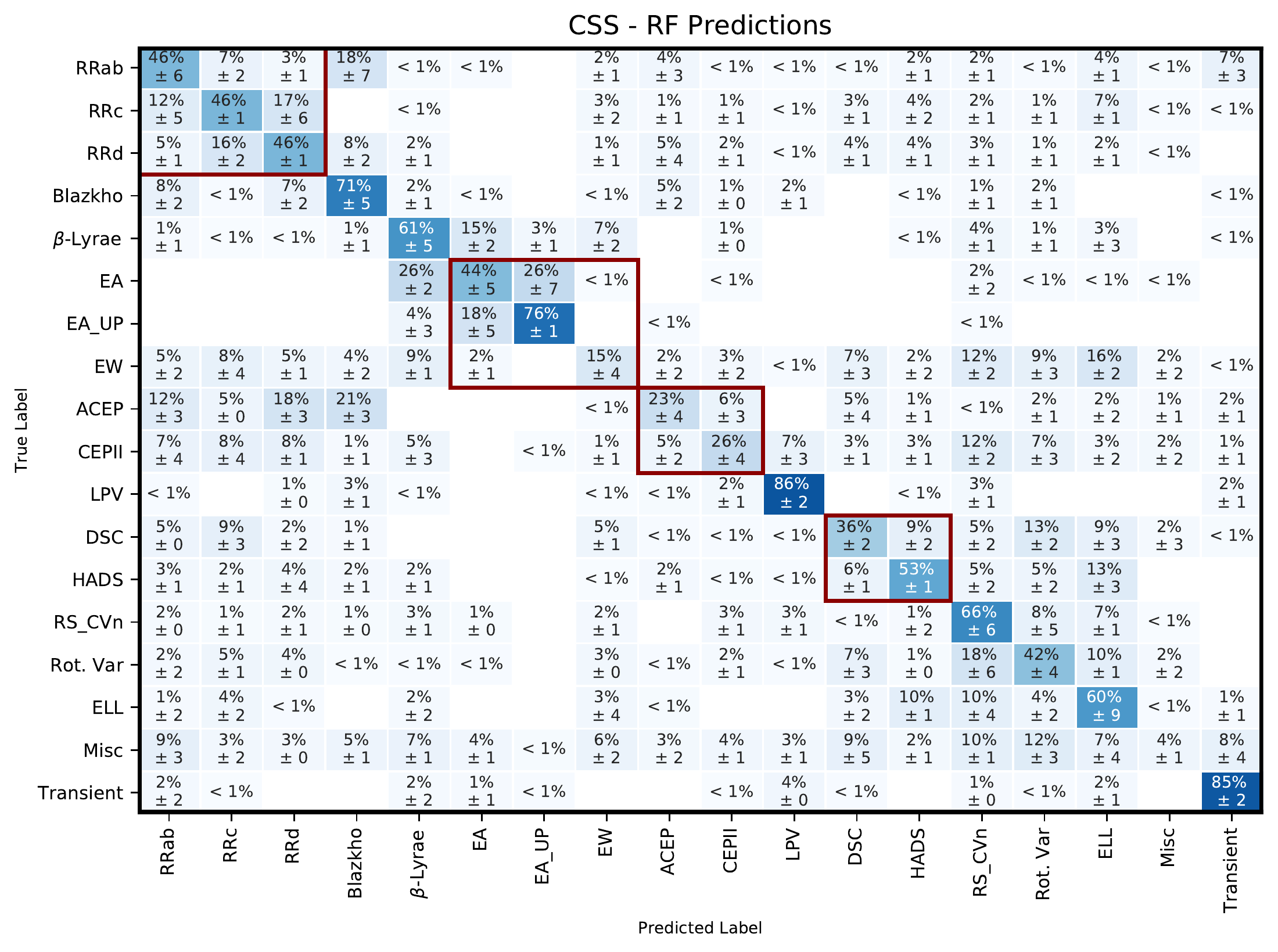}
    \caption{CSS confusion matrices for the predictions of the LSTM+PLSTM (L+P) and the Random Forest (RF) classifiers.}
    \label{cm:css}
\end{figure*}

This section presents and analyzes the confusion matrices associated with the best L+P and RF results, according to Table \ref{tab:summary}. Figures \ref{cm:linear}-\ref{cm:css} show the confusion matrices for the seven datasets we used in this work. Each percentage within the matrices indicates the classifier prediction as a fraction of the total number of true labels per class. Note that low dispersion in rows is related to a high precision score. Similarly, a high recall score is associated with low dispersion in columns.
\\\\
Confusion matrices show that the L+P model tends to be more precise than RF when predicting. RNNs learn representations of data to separate classes while the RF receives predefined features, standard for any time series. However, recurrent models depend strongly on the training set, which directly affects the representation's quality. For example, in Figure \ref{cm:wise}, the L+P can identify most of the ab-type RR Lyrae ($RRab$), but it is missing all c-type RRLyra ($RRc$), and most of the SX Phoenicis Delta Scuti (DSCT\_SXPHE).
\\\\
High precision scores often expose an overfitting problem related to class imbalance. In the example of Figure \ref{cm:wise}, the $RRab$ has 16412 samples versus the 3831 $RRc$ and 1098 DSCT\_SXPHE. It means that $RRab$ dominates training in terms of the number of samples, that the learned features will therefore be loaded towards this class. Though the network overfits $RRab$,  it can separate similar pulsating time-scales, such as $\delta$-scutis and RR Lyrae, because of their variability similitude. We use red boxes inside the matrices for grouping similar classes. We hypothesize that the significant variance in time scales (see Table \ref{tab:cadences}) affects the network to separate short-period objects. 

\section{Learning Curves}\label{appendix:lc}
\begin{figure*}
    \centering
    \includegraphics[scale=0.35]{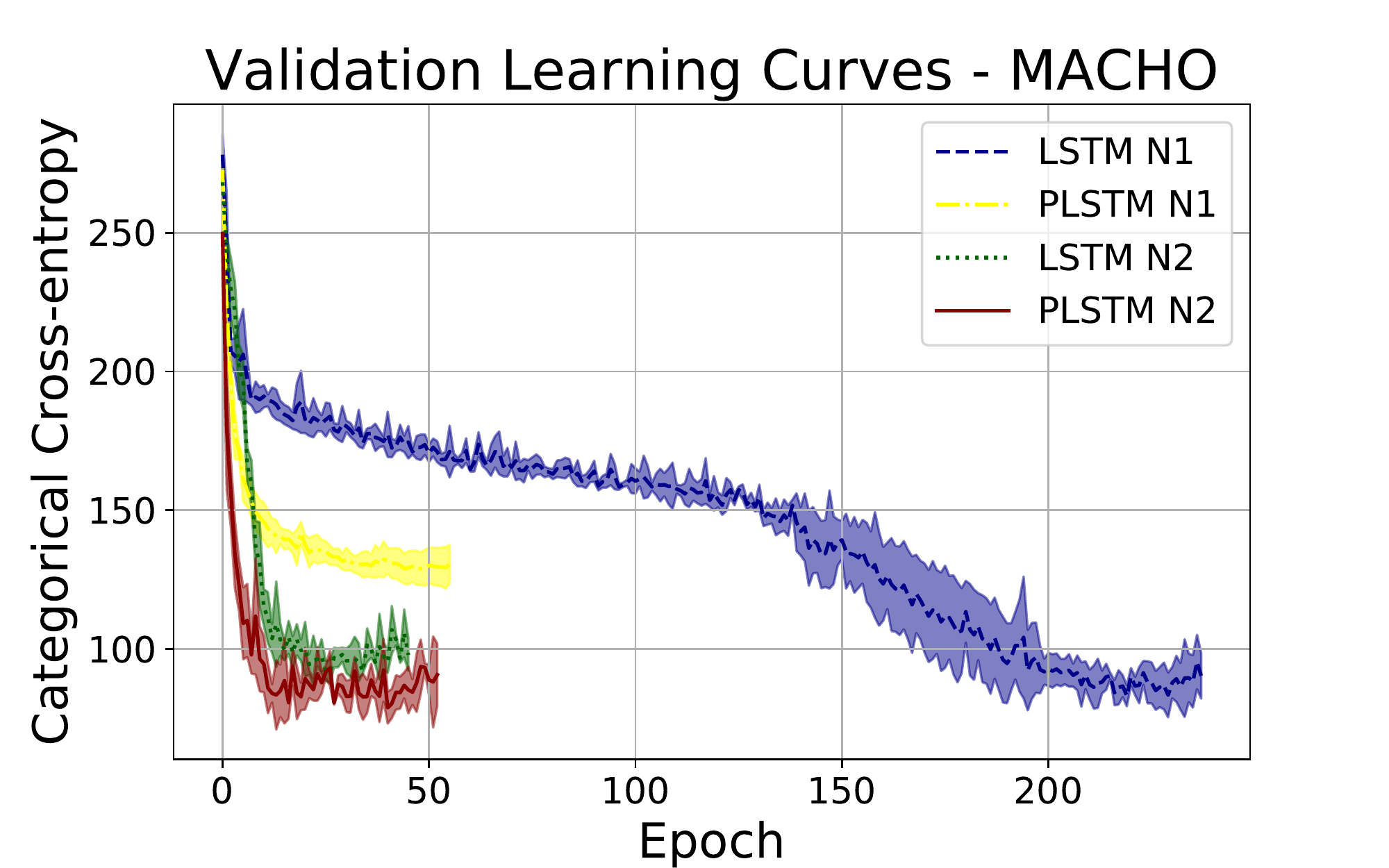}
    \includegraphics[scale=0.35]{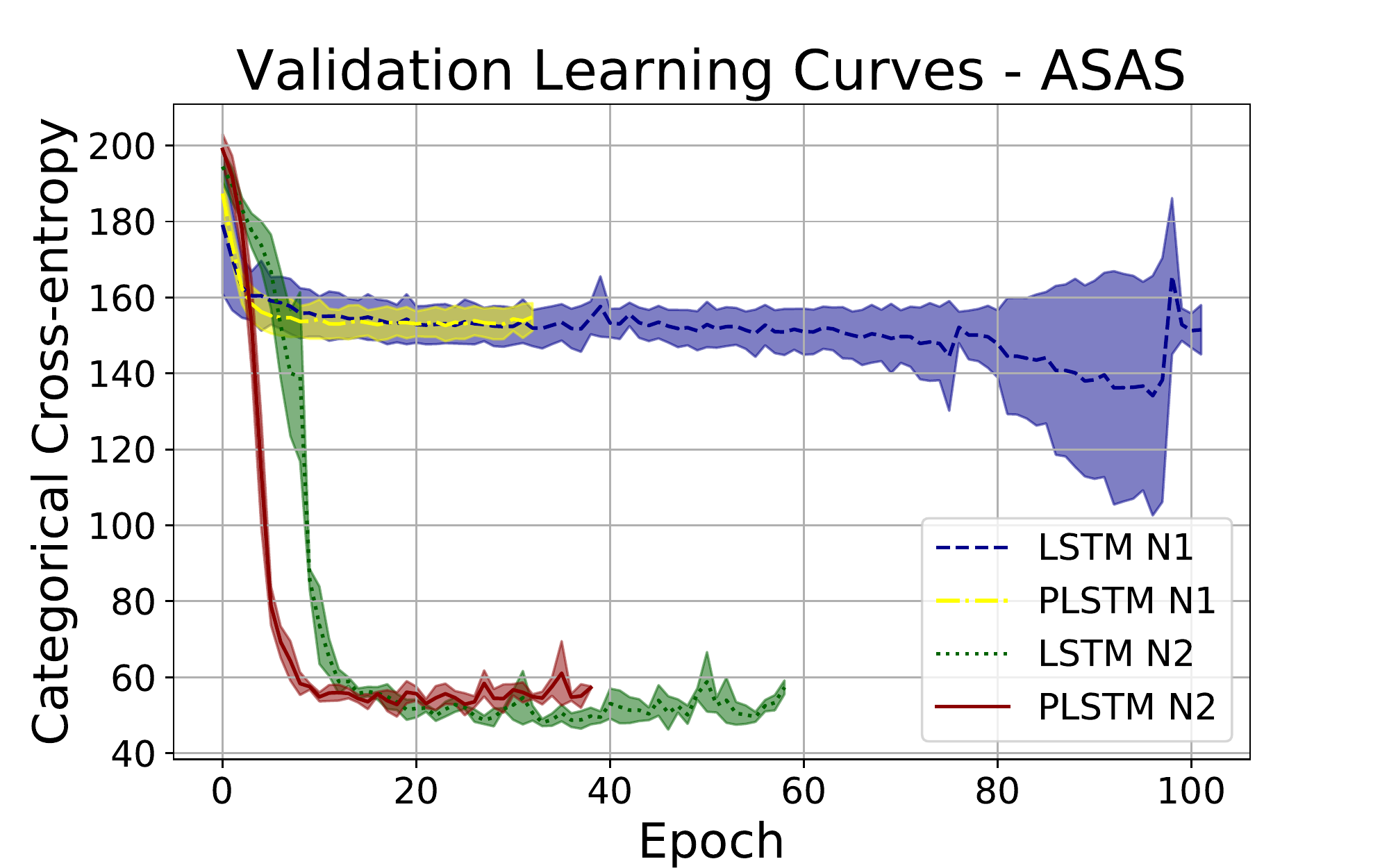}\\
    \includegraphics[scale=0.35]{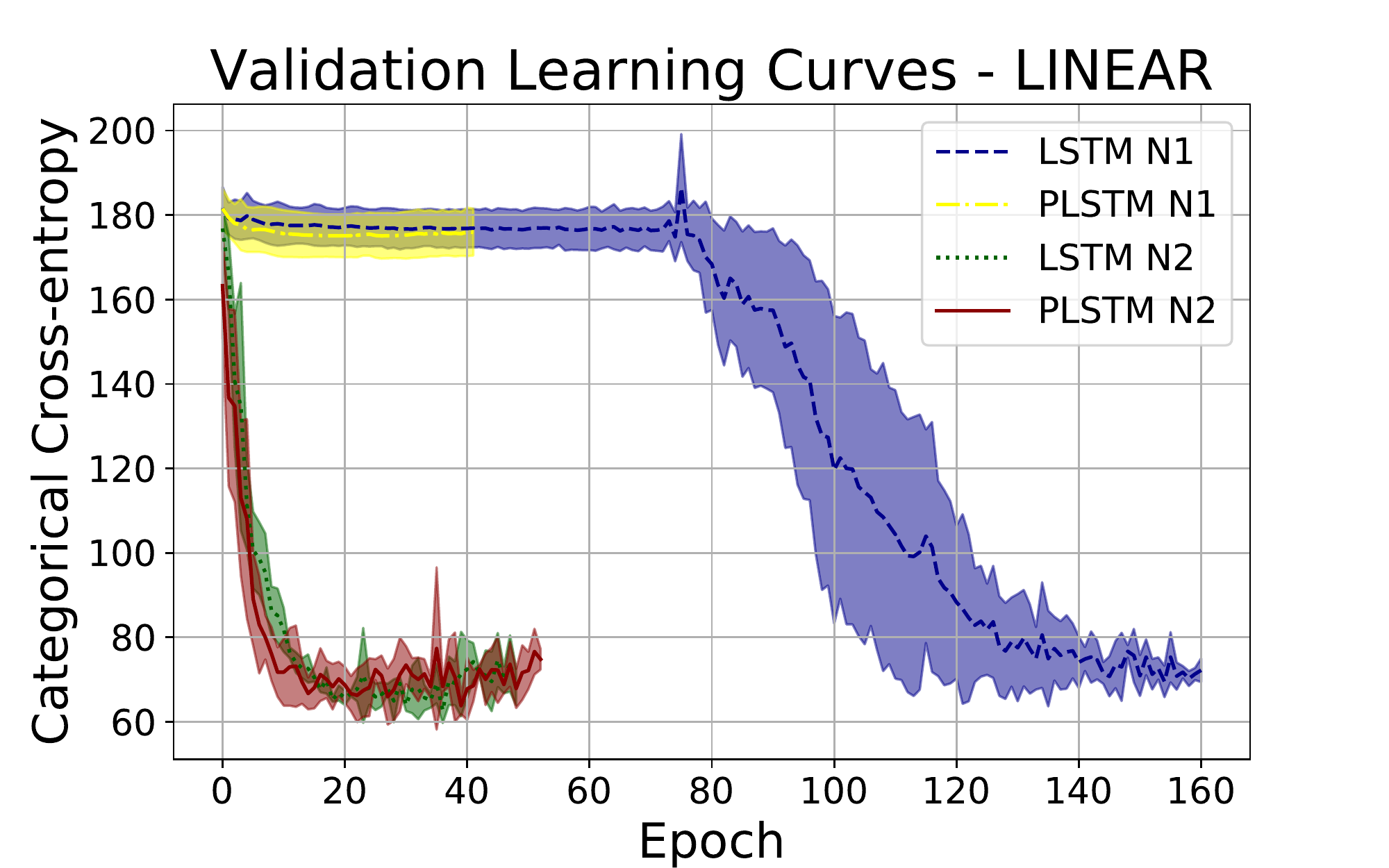}
    \includegraphics[scale=0.35]{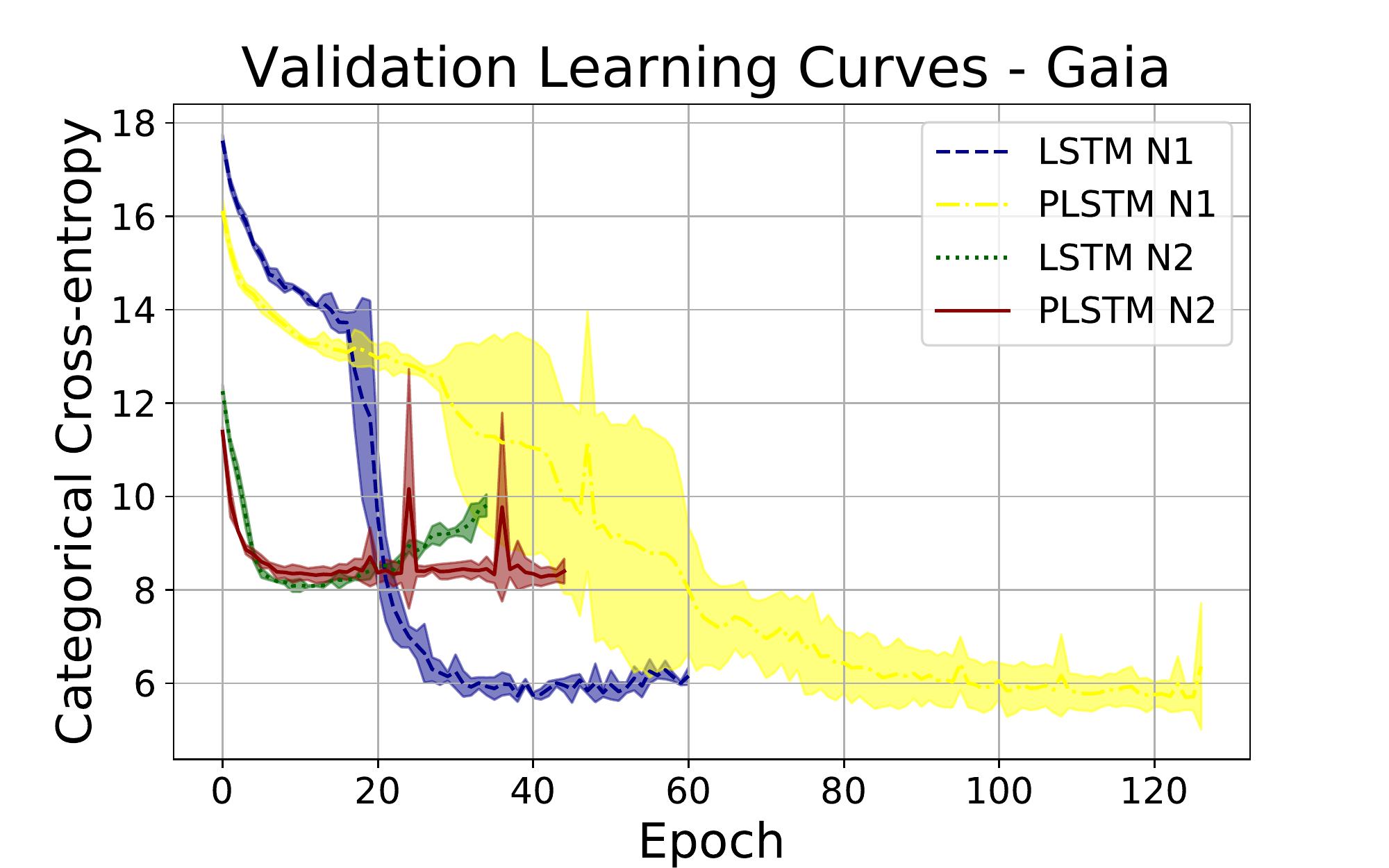}\\
    \includegraphics[scale=0.35]{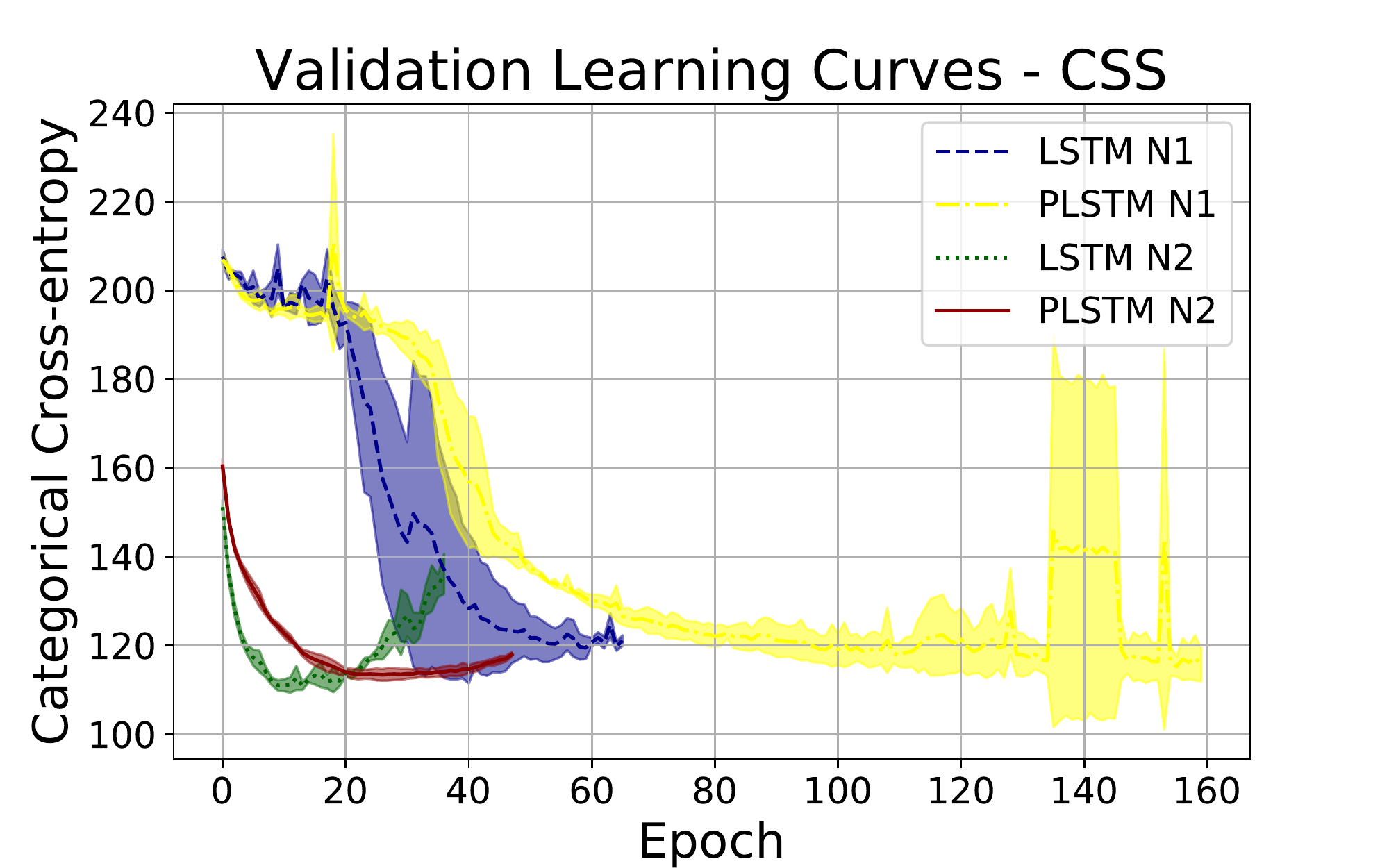}
    \includegraphics[scale=0.35]{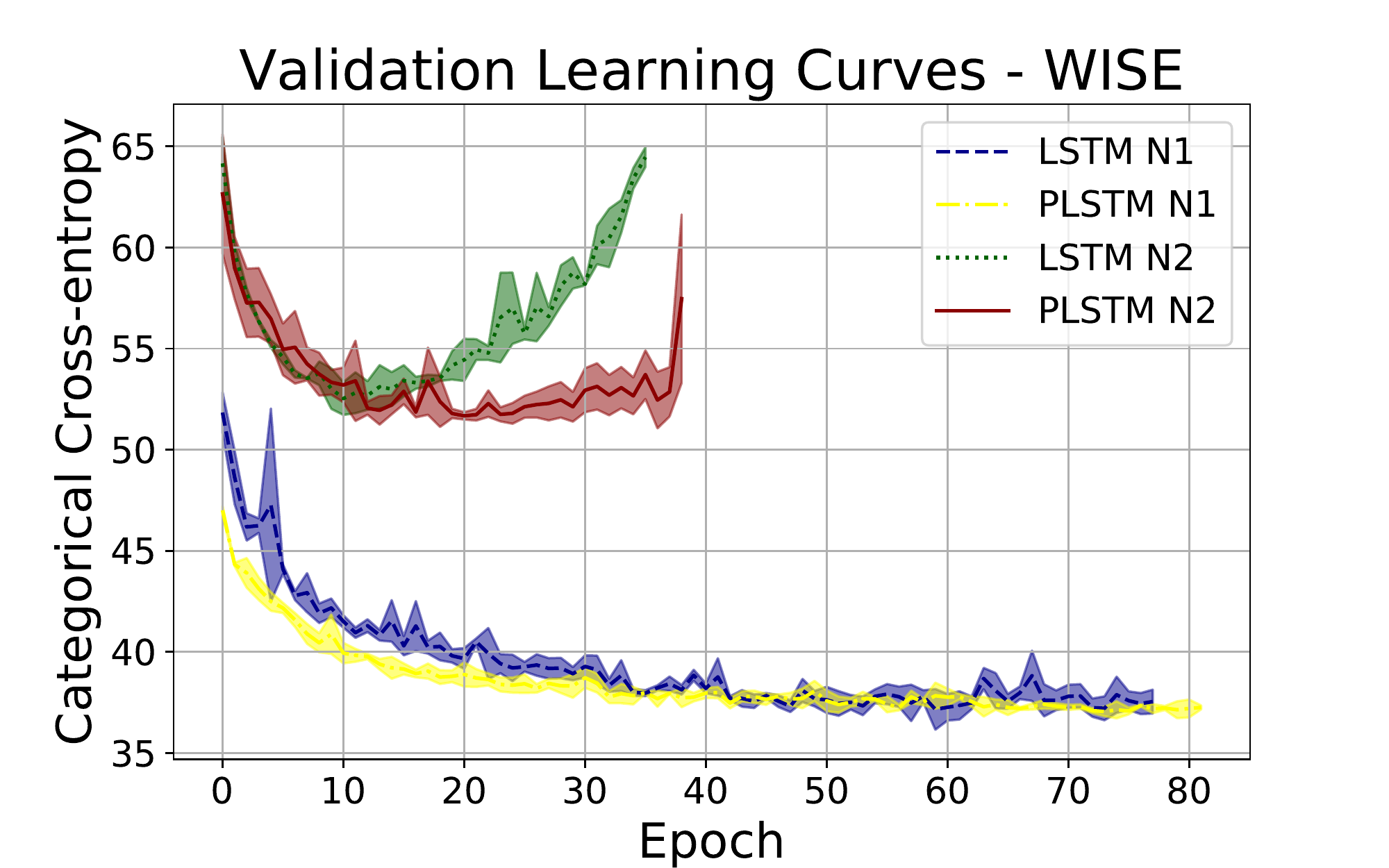}\\
    \includegraphics[scale=0.35]{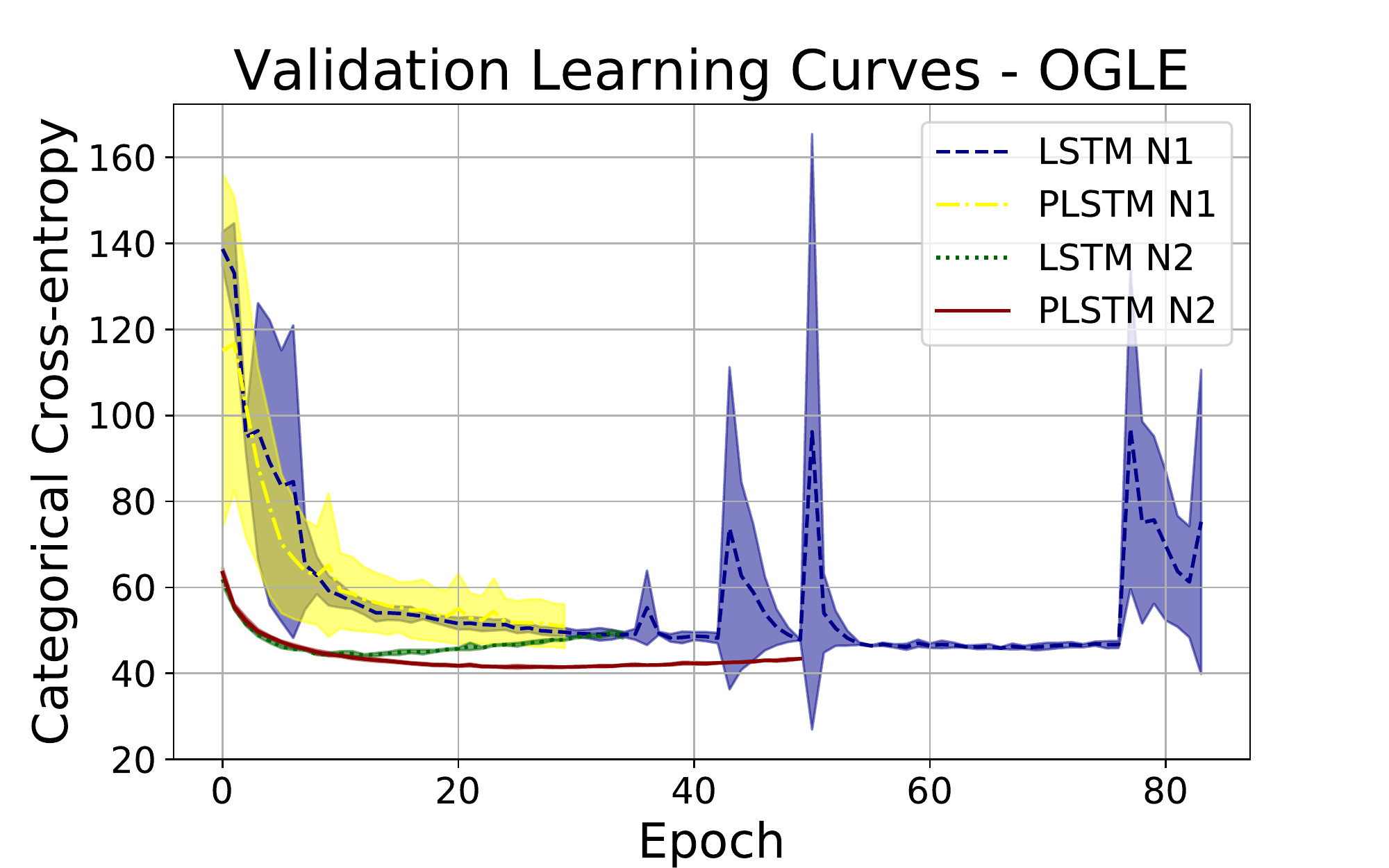}
    \caption{Validation Loss for the LSTM and PLSTM. The curves show the mean and standard deviation of the 3-fold execution. The input was normalized using min-max scaler (N1) or z-score standardization (N2)}
    \label{fig:lc}
\end{figure*}

Figure \ref{fig:lc} shows the learning curves corresponding to the LSTM and PLSTM, trained with min-max scaler (N1) and standardization (N2) as input normalization technique. Each plot presents the mean and standard deviation of the validation losses according to the cross-validation strategy explained in section \ref{modelselection}.
\\\\
In all datasets, the N2 normalization converged faster than the N1. However, it is not a guarantee of achieving the best validation loss. For example, in the WISE and Gaia dataset, the model trained on lightcurves N2 converged faster than N1 but with a worse validation loss.
\\\\
In general, the N2 normalization is similar to N1 in terms of minimum values but slower to converge. However, since the N2 normalization shift and scales lightcurves to have zero mean and unit standard deviation, the optimization becomes more manageable than in N1. We can see this effect in the variance of the learning curves in Figure \ref{appendix:lc}. We also tried using N1 on each lightcurve separately and N2 over the entire dataset. However, the results were not significant, and we empirically decided to use the highest scoring alternatives.

\section{FATS features}
Table \ref{tab:fatsfeatures} shows the features used to train the RF model. We used all single-band features on lightcurves with at least 10 observations.
\input{Tables/fats_features}


\bsp	
\label{lastpage}
\end{document}

%% file: Tables/data_tables.tex
\begin{table}
    \centering
    \begin{tabular}{c|c|}
        Label & Quantity \\ \hline
        Be & 128  \\
        CEPH & 101\\
        EB & 255  \\
        LPV & 365 \\
        MOA & 582 \\
        QSO & 59  \\
        RRL & 610 \\ \hline
        \textbf{Total}  &  \textbf{2100}  \\ 
    \end{tabular}
    \caption{MACHO dataset composition}
    \label{tab:macho_dataset}
\end{table}

\begin{table}
    \centering
    \begin{tabular}{c|c}
        Label & Quantity \\ \hline
        Beta\_Persei & 291  \\ 
        Delta\_Scuti & 70  \\ 
        RR\_Lyrae\_FM & 2234 \\ 
        RR\_Lyrae\_FO & 749 \\ 
        W\_Ursae\_Maj & 1860 \\ \hline
        \textbf{Total}  &  \textbf{5204}  \\
    \end{tabular}
    \caption{LINEAR dataset composition}
    \label{tab:linear_dataset}
\end{table}

\begin{table}
    \centering
    \begin{tabular}{c|c}
        Label & Quantity \\ \hline
        Beta Persei  &  349   \\ 
        Classical Cepheid  &  130   \\ 
        RR Lyrae FM  &  798   \\ 
        Semireg PV  &  184 \\ 
        W Ursae Maj  &  1639  \\ \hline
        \textbf{Total}  &  \textbf{3100}  \\ 
    \end{tabular}
    \caption{ASAS dataset composition}
    \label{tab:asas_dataset}
\end{table}

\begin{table}
    \centering
    \begin{tabular}{c|c}
        Label & Quantity \\ \hline
        EC & 6862 \\ 
        ED & 21503 \\ 
        ESD & 9475 \\ 
        Mira & 6090 \\ 
        OSARG & 234932 \\ 
        RRab & 25943  \\ 
        RRc & 7990  \\ 
        SRV & 34835  \\ 
        CEP & 7836  \\ 
        DSC & 2822 \\ 
        NonVar & 34815 \\ \hline
        \textbf{Total}  &  \textbf{393103}  \\ 
    \end{tabular}
    \caption{OGLE dataset composition}
    \label{tab:ogle_dataset}
\end{table}

\begin{table}
    \centering
    \begin{tabular}{c|c}
    Label & Quantity\\ \hline
    Blazkho & 243 \\ 
    CEPII & 277  \\ 
    DSC & 147  \\ 
    EA & 300  \\ 
    EA\_UP & 153  \\ 
    ELL & 142  \\ 
    EW & 300  \\ 
    HADS & 242  \\ 
    LPV & 300  \\ 
    Misc & 298  \\ 
    RRab & 300  \\ 
    RRc & 300  \\ 
    RRd & 300  \\ 
    RS\_CVn & 300  \\ 
    Rotational Var & 300  \\ 
    Transient & 300  \\ 
    beta\_Lyrae & 279  \\ \hline
    \textbf{Total}  &  \textbf{4481}  \\ 
    \end{tabular}
    \caption{CSS dataset composition}
    \label{tab:cata_dataset}
\end{table}

\begin{table}
    \centering
    \begin{tabular}{c|c}
    Label & Quantity\\ \hline
    CEP & 1884  \\
    DSCT\_SXPHE & 1098 \\
    Mira & 1396 \\
    NC & 2237 \\ 
    NonVar & 32795 \\ 
    OSARG & 53890 \\ 
    RRab & 16412 \\ 
    RRc & 3831 \\ 
    SRV & 8605 \\ \hline
    \textbf{Total}  &  \textbf{122148}  \\ 
    \end{tabular}
    \caption{WISE dataset composition}
    \label{tab:wise_dataset}
\end{table}

\begin{table}
    \centering
    \begin{tabular}{c|c}
    Label & Quantity \\ \hline
    CEP & 6484 \\ 
    DSCT\_SXPHE & 8579  \\ 
    MIRA\_SR & 150215  \\ 
    RRAB & 153392  \\ 
    RRC & 32206  \\ 
    RRD & 829  \\ 
    T2CEP & 1736  \\ \hline
    \textbf{Total}  &  \textbf{353441}  \\ 
    \end{tabular}
    \caption{Gaia dataset composition}
    \label{tab:gaia_dataset}
\end{table}

%% file: Tables/summary_table.tex
\begin{tabular}{|c|c|c|c|c|c|}
Dataset                  & Model                                        & Normalization                            & F1 Score                                       & Recall                                         & Precision                                      \\ \hline
                         & LSTM                                         & N1                                       & 0.881 $\pm$ 0.008                                 & 0.879 $\pm$ 0.006                                 & 0.878 $\pm$ 0.003                                 \\
                         & PLSTM                                        & N2                                       & 0.765 $\pm$ 0.034                                 & 0.803 $\pm$ 0.010                                 & 0.777 $\pm$ 0.023                                 \\
                         & {\color[HTML]{010066} \textbf{LSTM + PLSTM}} & {\color[HTML]{010066} \textbf{N1 \& N2}} & {\color[HTML]{010066} \textbf{0.881 $\pm$ 0.005}} & {\color[HTML]{010066} \textbf{0.881 $\pm$ 0.004}} & {\color[HTML]{010066} \textbf{0.879 $\pm$ 0.005}} \\
                         & RF                                           & N1                                       & 0.799 $\pm$ 0.025                                 & 0.842 $\pm$ 0.030                                 & 0.780 $\pm$ 0.017                                 \\
\multirow{-5}{*}{OGLE}   & \multicolumn{1}{l|}{Becker et.al., 2020}  & \multicolumn{1}{l|}{}                    & \multicolumn{1}{l|}{0.737 $\pm$ 0.005}            & \multicolumn{1}{l|}{0.730 $\pm$ 0.005}            & \multicolumn{1}{l|}{0.777 $\pm$ 0.006}            \\ \hline
                         & LSTM                                         & N1                                       & 0.598 $\pm$ 0.009                                 & 0.531 $\pm$ 0.008                                 & 0.543 $\pm$ 0.004                                 \\
                         & PLSTM                                        & N1                                       & 0.585 $\pm$ 0.006                                 & 0.529 $\pm$ 0.006                                 & 0.541 $\pm$ 0.006                                 \\
                         & {\color[HTML]{010066} \textbf{LSTM + PLSTM}} & {\color[HTML]{010066} \textbf{N1 \& N1}} & {\color[HTML]{010066} \textbf{0.646 $\pm$ 0.034}} & {\color[HTML]{333333} 0.537 $\pm$ 0.003}          & {\color[HTML]{010066} \textbf{0.551 $\pm$ 0.002}} \\
                         & RF                                           & N1                                       & 0.472 $\pm$ 0.006                                 & {\color[HTML]{010066} \textbf{0.563 $\pm$ 0.006}} & 0.463 $\pm$ 0.007                                 \\
\multirow{-5}{*}{WISE}   & \multicolumn{1}{l|}{Becker et.al., 2020}  & \multicolumn{1}{l|}{}                    & \multicolumn{1}{l|}{0.462 $\pm$ 0.004}            & \multicolumn{1}{l|}{0.450 $\pm$ 0.010}            & \multicolumn{1}{l|}{0.551 $\pm$ 0.055}            \\ \hline
                         & LSTM                                         & N1                                       & 0.805 $\pm$ 0.016                                 & 0.734 $\pm$ 0.022                                 & 0.760 $\pm$ 0.016                                 \\
                         & PLSTM                                        & N1                                       & 0.802 $\pm$ 0.027                                 & 0.733 $\pm$ 0.006                                 & 0.754 $\pm$ 0.003                                 \\
                         & {\color[HTML]{010066} \textbf{LSTM + PLSTM}} & {\color[HTML]{010066} \textbf{N1 \& N1}} & {\color[HTML]{010066} \textbf{0.863 $\pm$ 0.040}} & {\color[HTML]{333333} 0.741 $\pm$ 0.009}          & {\color[HTML]{010066} \textbf{0.772 $\pm$ 0.010}} \\
                         & RF                                           & N1                                       & 0.591 $\pm$ 0.000                                 & {\color[HTML]{010066} \textbf{0.814 $\pm$ 0.004}} & 0.569 $\pm$ 0.000                                 \\
\multirow{-5}{*}{GAIA}   & \multicolumn{1}{l|}{Becker et.al., 2020}  & \multicolumn{1}{l|}{}                    & \multicolumn{1}{l|}{0.668 $\pm$ 0.004}            & \multicolumn{1}{l|}{0.657 $\pm$ 0.006}            & \multicolumn{1}{l|}{0.713 $\pm$ 0.007}            \\ \hline
                         & LSTM                                         & N2                                       & 0.858 $\pm$ 0.016                                 & 0.848 $\pm$ 0.019                                 & {\color[HTML]{010066} \textbf{0.851 $\pm$ 0.003}} \\
                         & PLSTM                                        & N2                                       & 0.764 $\pm$ 0.026                                 & 0.763 $\pm$ 0.042                                 & 0.758 $\pm$ 0.030                                 \\
                         & {\color[HTML]{010066} \textbf{LSTM + PLSTM}} & {\color[HTML]{010066} \textbf{N2 \& N1}} & {\color[HTML]{010066} \textbf{0.877 $\pm$ 0.021}} & {\color[HTML]{333333} 0.822 $\pm$ 0.023}          & {\color[HTML]{333333} 0.844 $\pm$ 0.008}          \\
\multirow{-4}{*}{LINEAR} & RF                                           & N1                                       & 0.736 $\pm$ 0.007                                 & {\color[HTML]{010066} \textbf{0.850 $\pm$ 0.012}} & 0.687 $\pm$ 0.008                                 \\ \hline
                         & LSTM                                         & N2                                       & 0.836 $\pm$ 0.037                                 & 0.814 $\pm$ 0.012                                 & 0.813 $\pm$ 0.019                                 \\
                         & PLSTM                                        & N2                                       & 0.755 $\pm$ 0.032                                 & 0.771 $\pm$ 0.014                                 & 0.745 $\pm$ 0.018                                 \\
                         & {\color[HTML]{010066} \textbf{LSTM + PLSTM}} & {\color[HTML]{010066} \textbf{N2 \& N1}} & {\color[HTML]{010066} \textbf{0.921 $\pm$ 0.017}} & {\color[HTML]{010066} \textbf{0.867 $\pm$ 0.008}} & {\color[HTML]{010066} \textbf{0.886 $\pm$ 0.015}} \\
\multirow{-4}{*}{MACHO}  & RF                                           & N1                                       & 0.787 $\pm$ 0.082                                 & 0.818 $\pm$ 0.087                                 & 0.772 $\pm$ 0.078                                 \\ \hline
                         & LSTM                                         & N2                                       & 0.500 $\pm$ 0.027                                 & 0.506 $\pm$ 0.011                                 & 0.490 $\pm$ 0.016                                 \\
                         & PLSTM                                        & N1                                       & 0.396 $\pm$ 0.008                                 & 0.403 $\pm$ 0.016                                 & 0.386 $\pm$ 0.013                                 \\
                         & {\color[HTML]{010066} \textbf{LSTM + PLSTM}} & {\color[HTML]{010066} \textbf{N2 \& N1}} & {\color[HTML]{010066} \textbf{0.505 $\pm$ 0.030}} & {\color[HTML]{010066} \textbf{0.519 $\pm$ 0.019}} & {\color[HTML]{010066} \textbf{0.496 $\pm$ 0.020}} \\
\multirow{-4}{*}{CSS}    & RF                                           & N1                                       & 0.464 $\pm$ 0.004                                 & 0.493 $\pm$ 0.004                                 & 0.474 $\pm$ 0.006                                 \\ \hline
                         & LSTM                                         & N2                                       & 0.931 $\pm$ 0.005                                 & 0.902 $\pm$ 0.027                                 & 0.911 $\pm$ 0.017                                 \\
                         & PLSTM                                        & N2                                       & 0.892 $\pm$ 0.020                                 & 0.894 $\pm$ 0.016                                 & 0.891 $\pm$ 0.014                                 \\
                         & {\color[HTML]{010066} \textbf{LSTM + PLSTM}} & {\color[HTML]{010066} \textbf{N2 \& N1}} & {\color[HTML]{010066} \textbf{0.957 $\pm$ 0.007}} & {\color[HTML]{333333} 0.920 $\pm$ 0.015}          & {\color[HTML]{010066} \textbf{0.935 $\pm$ 0.011}} \\
\multirow{-4}{*}{ASAS}   & RF                                           & N1                                       & 0.917 $\pm$ 0.023                                 & {\color[HTML]{010066} \textbf{0.961 $\pm$ 0.012}} & 0.891 $\pm$ 0.027                                 \\ \hline
\end{tabular}

%% file: Tables/fats_features.tex
\begin{center}
\begin{table}
    \centering
    \begin{tabular}{|c|c|}
    \hline
    \textbf{FATS Feature Name}               & \textbf{Reference}   \\ \hline\hline
    MedianAbsDev                    & \cite{richards2011machine} \\ \hline
    PeriodLS                        & \cite{kim2011quasi}     \\ \hline
    Autocor\_length                 & \cite{kim2011quasi}     \\ \hline
    Q31                             & \cite{kim2014epoch}     \\ \hline
    FluxPercentileRatioMid35        & \cite{richards2011machine} \\ \hline
    FluxPercentileRatioMid50        & \cite{richards2011machine} \\ \hline
    FluxPercentileRatioMid65        & \cite{richards2011machine} \\ \hline
    FluxPercentileRatioMid20        & \cite{richards2011machine} \\ \hline
    Beyond1Std                      & \cite{richards2011machine} \\ \hline
    Skew                            & \cite{richards2011machine} \\ \hline
    Con                             & \cite{kim2014epoch}     \\ \hline
    Gskew                           & \cite{nun2015fats}  \\ \hline
    Meanvariance                    & \cite{kim2011quasi}     \\ \hline
    StetsonK                        & \cite{richards2011machine} \\ \hline
    FluxPercentileRatioMid80        & \cite{richards2011machine} \\ \hline
    CAR\_sigma                      & \cite{pichara2012improved} \\ \hline
    CAR\_mean                       & \cite{pichara2012improved} \\ \hline
    CAR\_tau                        & \cite{pichara2012improved} \\ \hline
    MedianBRP                       & \cite{richards2011machine} \\ \hline
    Rcs                             & \cite{kim2011quasi}     \\ \hline
    Std                             & \cite{richards2011machine} \\ \hline
    SmallKurtosis                   & \cite{richards2011machine} \\ \hline
    Mean                            & \cite{kim2014epoch}     \\ \hline
    Amplitude                       & \cite{richards2011machine} \\ \hline
    StructureFunction\_index\_21    & \cite{nun2015fats} \\ \hline
    StructureFunction\_index\_31    & \cite{nun2015fats} \\ \hline
    Freq1\_harmonics\_amplitude\_0  & \cite{richards2011machine} \\ \hline
    \end{tabular}
    \caption{FATS features used for Random Forest training}
    \label{tab:fatsfeatures}
\end{table}    
\end{center}

\begin{center}
\begin{table}
    \centering
    \begin{tabular}{|c|c|}
    \hline
    \textbf{FATS Feature Name}               & \textbf{Reference}   \\ \hline\hline
    Freq1\_harmonics\_amplitude\_1  & \cite{richards2011machine} \\ \hline
    Freq1\_harmonics\_amplitude\_2  & \cite{richards2011machine} \\ \hline
    Freq1\_harmonics\_amplitude\_3  & \cite{richards2011machine} \\ \hline
    Freq1\_harmonics\_rel\_phase\_0 & \cite{richards2011machine} \\ \hline
    Freq1\_harmonics\_rel\_phase\_1 & \cite{richards2011machine} \\ \hline
    Freq1\_harmonics\_rel\_phase\_2 & \cite{richards2011machine} \\ \hline
    Freq1\_harmonics\_rel\_phase\_3 & \cite{richards2011machine} \\ \hline
    Freq2\_harmonics\_amplitude\_0  & \cite{richards2011machine} \\ \hline
    Freq2\_harmonics\_amplitude\_1  & \cite{richards2011machine} \\ \hline
    Freq2\_harmonics\_amplitude\_2  & \cite{richards2011machine} \\ \hline
    Freq2\_harmonics\_amplitude\_3  & \cite{richards2011machine} \\ \hline
    Freq2\_harmonics\_rel\_phase\_0 & \cite{richards2011machine} \\ \hline
    Freq2\_harmonics\_rel\_phase\_1 & \cite{richards2011machine} \\ \hline
    Freq2\_harmonics\_rel\_phase\_2 & \cite{richards2011machine} \\ \hline
    Freq2\_harmonics\_rel\_phase\_3 & \cite{richards2011machine} \\ \hline
    Freq3\_harmonics\_amplitude\_0  & \cite{richards2011machine} \\ \hline
    Freq3\_harmonics\_amplitude\_1  & \cite{richards2011machine} \\ \hline
    Freq3\_harmonics\_amplitude\_2  & \cite{richards2011machine} \\ \hline
    Freq3\_harmonics\_amplitude\_3  & \cite{richards2011machine} \\ \hline
    Freq3\_harmonics\_rel\_phase\_0 & \cite{richards2011machine} \\ \hline
    Freq3\_harmonics\_rel\_phase\_1 & \cite{richards2011machine} \\ \hline
    Freq3\_harmonics\_rel\_phase\_2 & \cite{richards2011machine} \\ \hline
    Freq3\_harmonics\_rel\_phase\_3 & \cite{richards2011machine} \\ \hline
    Psi\_CS                         & \cite{kim2014epoch}     \\ \hline
    Psi\_eta                        & \cite{kim2014epoch}     \\ \hline
    AndersonDarling                 & \cite{kim2009detrending}     \\ \hline
    LinearTrend                     & \cite{richards2011machine} \\ \hline
    PercentAmplitude                & \cite{richards2011machine} \\ \hline
    MaxSlope                        & \cite{richards2011machine} \\ \hline
    PairSlopeTrend                  & \cite{richards2011machine} \\ \hline
    Period\_fit                     & \cite{kim2011quasi}     \\ \hline
    StructureFunction\_index\_32    &  \cite{nun2015fats}    \\ \hline
    \end{tabular}
\end{table}    
\end{center}